%% file: main.tex
\documentclass[10pt,journal,compsoc]{IEEEtran}
\ifCLASSOPTIONcompsoc
  \usepackage[nocompress]{cite}
\else
  \usepackage{cite}
\fi
\ifCLASSINFOpdf
\else
\fi
\usepackage{microtype}                 
\PassOptionsToPackage{warn}{textcomp}  
\usepackage{textcomp}                  
\usepackage{mathptmx}                  
\usepackage{times}                     
\usepackage{cite}                      
\usepackage{tabu}                      
\usepackage{booktabs}                  

\usepackage[english]{babel} 
\usepackage{blindtext}
\usepackage{amsmath}
\usepackage{amssymb}
\usepackage{mathtools}
\usepackage{diagbox}
\usepackage{multirow}
\usepackage{diagbox}
\usepackage{tikz}
\usepackage{makecell}
\usepackage{supertabular}
\usepackage{longtable}
\usepackage{ulem}
\usepackage{url}
\usepackage{tcolorbox}
\usepackage{soul}
\usepackage{hyperref}

\usepackage{lipsum}

\newcommand{\textbfsub}[2]{\textbf{#1\textsubscript{#2}}}

\newcommand\alexout{\bgroup\markoverwith{\textcolor{red}{\rule[0.5ex]{2pt}{0.4pt}}}\ULon}

\newcommand{\alexin}[1]{{\textcolor[HTML]{000000}{#1}}}
\definecolor{subcomponent}{RGB}{255, 255, 255}
\newcommand*\subcomponent[1]{\tikz[baseline=(char.base)]{
            \node[shape=circle,fill=subcomponent,text=black, inner sep= 0.5pt,draw=black] (char) {\textsf{\small #1}}}}

\definecolor{component}{RGB}{0, 0, 0}        
\newcommand*\component[1]{\tikz[baseline=(char.base)]{
            \node[shape=circle,fill=component,text=white, inner sep= 0.5pt,draw=black] (char) {\textsf{\small #1}}}}
\newcommand{\tool}{\textit{AdaVis}}

\definecolor{SF}{HTML}{FFE38C}
\newcommand{\SF}[1]{\tcbox[on line, boxsep=1pt, left=0pt,right=0pt,top=0pt,bottom=0pt, colframe=white,colback=SF] {#1}}

\definecolor{CF}{HTML}{F4B183}
\newcommand{\CF}[1]{\tcbox[on line, boxsep=1pt, left=0pt,right=0pt,top=0pt,bottom=0pt, colframe=white,colback=CF] {#1}}

\definecolor{COL}{HTML}{ECECEC}
\newcommand{\COL}[1]{\tcbox[on line, boxsep=1pt, left=0pt,right=0pt,top=0pt,bottom=0pt, colframe=white,colback=COL] {#1}}

\definecolor{DS}{HTML}{CCB299}
\newcommand{\DS}[1]{\tcbox[on line, boxsep=1pt, left=0pt,right=0pt,top=0pt,bottom=0pt, colframe=white,colback=DS] {#1}}

\definecolor{VIS}{HTML}{CDE9BB}
\newcommand{\VIS}[1]{\tcbox[on line, boxsep=1pt, left=0pt,right=0pt,top=0pt,bottom=0pt, colframe=white,colback=VIS] {#1}}

\begin{document}
%
\title{\tool{}: Adaptive and Explainable Visualization Recommendation for Tabular Data}
%
%
%
%

\author{Songheng~Zhang,
        Haotian~Li,
        ~Huamin~Qu and Yong~Wang
\IEEEcompsocitemizethanks{\IEEEcompsocthanksitem S. Zhang and Y. Wang are with the School
of Computing and Information Systems, Singapore Management University, Singapore.
Y. Wang is the corresponding author.
E-mail: \{shzhang.2021, yongwang\}@smu.edu.sg.
\IEEEcompsocthanksitem H. Li and H. Qu are with the Hong Kong University of Science and Technology
E-mail: haotian.li@connect.ust.hk, huamin@cse.ust.hk. This work was done when Haotian Li was a visiting student supervised by Dr. Yong Wang at Singapore Management University.

}

}

%
%

\markboth{IEEE TRANSACTIONS ON VISUALIZATION AND COMPUTER GRAPHICS, VOL. XX, NO. XX, XX 2018}%
{Shell \MakeLowercase{\textit{et al.}}: Bare Demo of IEEEtran.cls for Computer Society Journals}
%



\IEEEtitleabstractindextext{%
\begin{abstract}
Automated visualization recommendation facilitates the rapid creation of effective visualizations, which is especially beneficial for users with limited time and limited knowledge of data visualization.
There is an increasing trend in leveraging machine learning (ML) techniques to achieve an end-to-end visualization recommendation.
However, existing ML-based approaches implicitly assume that there is only one appropriate visualization for a specific dataset, which is often not true for real applications. Also, they often work like a black box, and are difficult for users to understand the reasons for recommending specific visualizations.
To fill the research gap, we propose \tool{}, an adaptive and explainable approach to recommend one or multiple appropriate visualizations for a tabular dataset.
It leverages a box embedding-based knowledge graph to well model the possible one-to-many mapping relations among different
entities (i.e., data features, dataset columns, datasets, and visualization choices).
The embeddings of the entities and relations can be learned from dataset-visualization pairs.
Also, \tool{} incorporates the attention mechanism into the inference framework. Attention can indicate the relative importance of data features for a dataset and provide fine-grained explainability.
Our extensive evaluations through quantitative metric evaluations, case studies, and \alexin{user} interviews demonstrate the effectiveness of \tool{}.
\end{abstract}

}

\maketitle

\IEEEdisplaynontitleabstractindextext

%
\IEEEpeerreviewmaketitle

\input{src/1-intro}

\input{src/2-relatedwork}

\input{src/3-method}

\input{src/4-evaluation}
\input{src/5-discussion}

\input{src/6-conclusion}

\ifCLASSOPTIONcompsoc
  \section*{Acknowledgments}
\else
  \section*{Acknowledgment}
\fi
This project is supported by the Ministry of Education, Singapore, under its Academic Research Fund Tier 2 (Proposal ID: T2EP20222-0049) and HK RGC GRF grant 16210722. Any opinions, findings and conclusions, or recommendations expressed in this material are those of the author(s) and do not reflect the views of the Ministry of Education, Singapore.
We are grateful to 
Xiaolin Wen for his help in figure editing, to the experts in participating our interviews, and to anonymous reviewers for their constructive feedback.

\ifCLASSOPTIONcaptionsoff
  \newpage
\fi



\bibliographystyle{IEEEtran}
\bibliography{main}
\input{src/appendix.tex}

\end{document}

%% file: src/1-intro.tex
\IEEEraisesectionheading{\section{Introduction}\label{sec:intro}}
 \maketitle

\IEEEPARstart{D}{ata} visualization
has become increasingly popular in
data analytics and insight communication.
It is common to create visualizations for tabular
datasets in various domains, including investment, sales, engineering, education, and scientific research~\cite{zhou2020table2charts,ward2010interactive,lin2022dminer}.
However, creating compelling visualizations
requires expertise in data visualization and relies on manual specifications through either programming or mouse interactions (e.g., clicking and dragging, and dropping).
The visualization tools can generally be categorized into two types:
visualization packages (e.g., ggplot2~\cite{wickham2010layered}, Vega~\cite{satyanarayan2015reactive}, D3~\cite{bostock2011d3}, and Prefuse~\cite{heer2005prefuse}) and visualization software (e.g., Tableau~\footnote{\url{https://www.tableau.com/}} and Microsoft Power BI~\footnote{\url{https://powerbi.microsoft.com/en-us/desktop/}}). The former needs users to do programming with different languages (e.g., Python, R, Java, and JavaScript), and the latter often asks users to manually drag and drop and specify the mapping between data and visual encodings.
As a result, it is often complicated and time-consuming for common users without a background in data visualization to generate effective visualizations.

Automatic visualization recommendation techniques have been proposed to make the visualization creation process more accessible and more efficient~\cite{dibia2019data2vis,wang2019deepdrawing,hu2019vizml,li2022diverse, Zeng2021AnEF}. Among them, the machine learning (ML) based visualization recommendation approaches have become popular
in the past few years.
They
are often data-driven and implicitly model the mapping between datasets
and appropriate visualizations.
Compared with other rule-based visualization recommendation techniques (e.g., APT~\cite{apt1986machinlay}, Show Me~\cite{mackinlay2007show} and SeeDB~\cite{vartak2014seedb}), the ML-based approaches have the advantage of being an end-to-end visualization recommendation without specifying heuristics, 
and there have been an increasing number of research studies to achieve better visualization recommendations by leveraging machine learning techniques~\cite{zhu2020survey,wang2021survey,wu2021ai4vis,chen2019towards}. 
\begin{figure}[!ht]
    \centering
    \includegraphics[width=\linewidth]{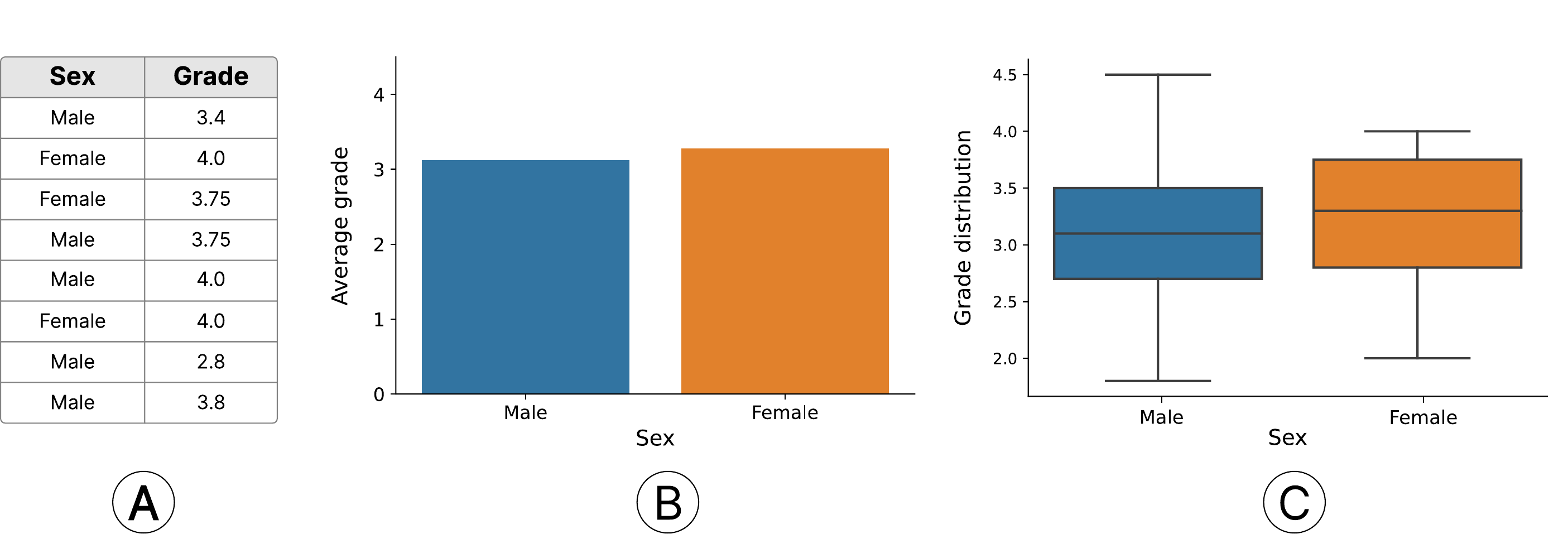}
    \caption{A dataset regarding students' grades in a class. The dataset can be visualized in a bar chart or a box plot.}
    \vspace{-1em}
    \label{fig: gpa}
\end{figure}

However, existing ML-based visualization recommendation approaches suffer from two major issues:
\textit{\textbf{adaptability}} and \textit{\textbf{explainability}}.
First, the ML-based approaches implicitly assume a one-to-one correspondence between datasets and visualizations by training on dataset-visualization pairs.
However, it does not always hold in real applications.
For example, as shown in Figure~\ref{fig: gpa},
the dataset that contains students' grades over a course can be visualized as either a bar chart showing the average grades of female and male students or a boxplot displaying the grades distribution of female and male students.
Both visualizations are suitable in terms of visual encodings and the optimal choice can be either of them, depending on users' preferred level of details of GPA distribution.

\begin{figure*}[ht!]
  \centering\includegraphics[width=\linewidth]{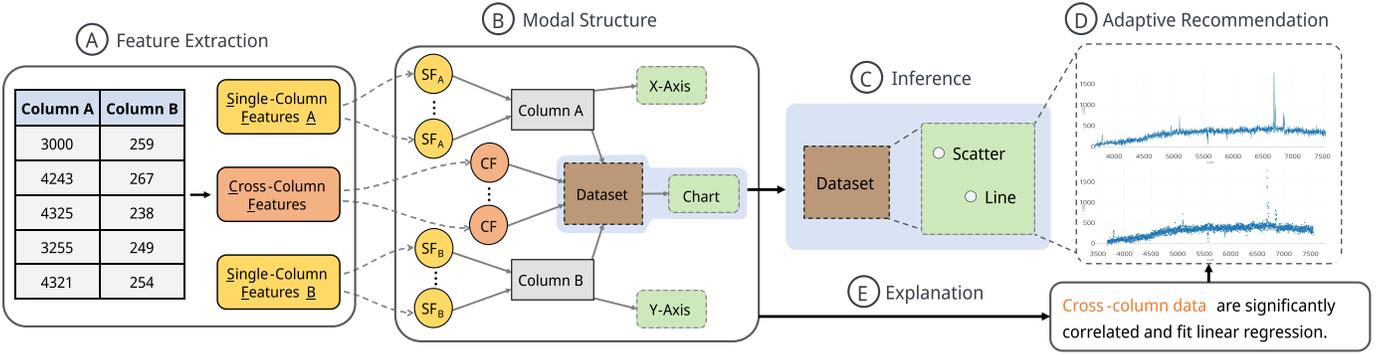}
  \caption{The workflow of \tool{} recommendation consists of Feature Extraction, Model Structure, Inference, Explanation Generation and Adaptive Recommendation. \protect\subcomponent{A} delineates that \tool{} extracts cross-column and single-column features from a dataset; \protect\subcomponent{B} displays the structure of the \tool{}: it uses extracted features to infer appropriate visualization choices; \protect\subcomponent{C} illustrates the inference details: the visualization types within a box are identified as visualization type recommendations for the dataset; \protect\subcomponent{D} displays the multiple appropriate visualization types for the dataset; \protect\subcomponent{E} illustrates that \tool{} provides an explanation for the recommendation results.}
  \label{fig:teaser}
  \vspace{-2em}
\end{figure*}

According to our survey, no existing ML-based approaches can adaptively recommend multiple appropriate visualizations for a dataset. 
Existing ML-based approaches often recommend the only one visualization choice~\cite{li2021kg4vis,hu2019vizml,zhou2020table2charts, dibia2019data2vis},
not adaptive to different datasets.
%
%
Second, most ML-based
approaches (e.g., Data2Vis~\cite{dibia2019data2vis} and VizML~\cite{hu2019vizml}) are built upon deep neural networks.
For instance, a three-layer fully connected neural network is employed in VizML~\cite{hu2019vizml} to predict the axis encodings and visualization types for a dataset.
These approaches work as a black box and can undermine users' trust in the visualization recommendation results, especially for general users without a background in visualization and deep neural networks.
A recent study, KG4Vis~\cite{li2021kg4vis}, presents a knowledge-graph-based approach to recommend visualization in an explainable manner for tabular datasets.
But their explanations are built on single-column features and can only indicate which single-column features account more for a visualization type, which is called \textit{global interpretation}~\cite{murdoch2019definitions}. The global interpretation captures the overall relationship the model learned from all datasets, identifying which feature values are strongly associated with specific visualization types. However, KG4Vis neglects cross-column features and fails to provide a fine-grained explanation for each unique dataset. The \textit{local interpretation}~\cite{murdoch2019definitions}
that focuses on explaining a particular visualization recommendation for an individual dataset is still missing 

In this paper, we propose \tool{}, an \underline{\textbf{Ad}}aptive and Expl\underline{\textbf{a}}inable \underline{\textbf{Vis}}ualization Recommendation approach for tabular data
through logical reasoning over knowledge graphs.
Inspired by KG4Vis~\cite{li2021kg4vis}, our approach also leverages a knowledge graph to model the relations between different entities involved in visualization recommendation (Figure~\ref{fig:teaser} \protect\subcomponent{\textbf{A}}~\protect\subcomponent{\textbf{B}}), e.g., \textit{data features}, \textit{dataset columns}, \textit{datasets} and \textit{visualization design choices}.
The relations in the knowledge graph define the correspondence between two different types of entities. For example, ``(a dataset) is visualized by (a visualization choice)''.
Such relations intrinsically specify the inference rules in visualization designs.
However, instead of employing the widely-used vector embeddings~\cite{transe2013bordes, li2021kg4vis} to indicate the inference results, we adopt box embeddings~\cite{ren2020query2box} that essentially allow the visualization recommendation results to cover multiple appropriate visualization choices for a given dataset (Figure~\ref{fig:teaser}~\protect\subcomponent{\textbf{C}}).
The incorporation of box embeddings leads to better \textit{\textbf{adaptability}} for visualization recommendation, 
enabling \tool{} to
adaptively recommend 
an appropriate number of visualization choices based on the characteristics of a dataset. 
Also, we have incorporated an attention mechanism into \tool{}, which assesses the importance of different features for visualization recommendations~\cite{chen2021multi}. This mechanism works over the knowledge graph, ensuring that our recommendations (Figure~\ref{fig:teaser}~\protect\subcomponent{\textbf{D}}) are informed by relevant data features. 
Moreover, \tool{} offers fine-grained explanations for the visualization recommendations for a specific dataset (\textit{local interpretation}) by tracing the importance of data features along inference paths, thereby improving the interpretability of our recommendations.
The explanations (Figure~\ref{fig:teaser}~\protect\subcomponent{\textbf{E}}) are natural language (NL) sentences automatically generated from rule-based templates.

We extensively evaluated the effectiveness and usability of \tool{} by using the dataset-visualization pairs collected by Hu et al.~\cite{hu2019vizml}.
We first quantitatively compared \tool{} with other state-of-the-art baseline approaches
in terms of visualization recommendation accuracy.
Then, we showed a gallery of visualization recommendation results and the corresponding natural language explanations to demonstrate the adaptability and explainability of \tool{}. Further, we conducted user interviews to invite
both data visualization experts and common users to verify whether the recommended visualizations are meaningful and align well with their domain knowledge of visualization design requirements and whether explanations regarding these recommendations are correct.

The paper's main contributions can be summarized as follows:
\begin{itemize}
\setlength\itemsep{0pt}
    \item We propose \tool{}, an adaptive and explainable visualization recommendation approach for tabular data via knowledge graphs.
    It adaptively recommends multiple appropriate visualizations for a specific dataset, better modeling the real visualization design process. Also, it can provide fine-grained explanations for different datasets.
    \item We extensively assess \tool{} through quantitative metric comparisons with other baseline approaches, qualitative case studies, and user interviews.
    The results demonstrate the effectiveness and usability of \tool{} in providing adaptive and explainable visualization recommendations.

\end{itemize}

%% file: src/2-relatedwork.tex
\begin{figure}[!ht]
    \centering
    \includegraphics[width=\linewidth]{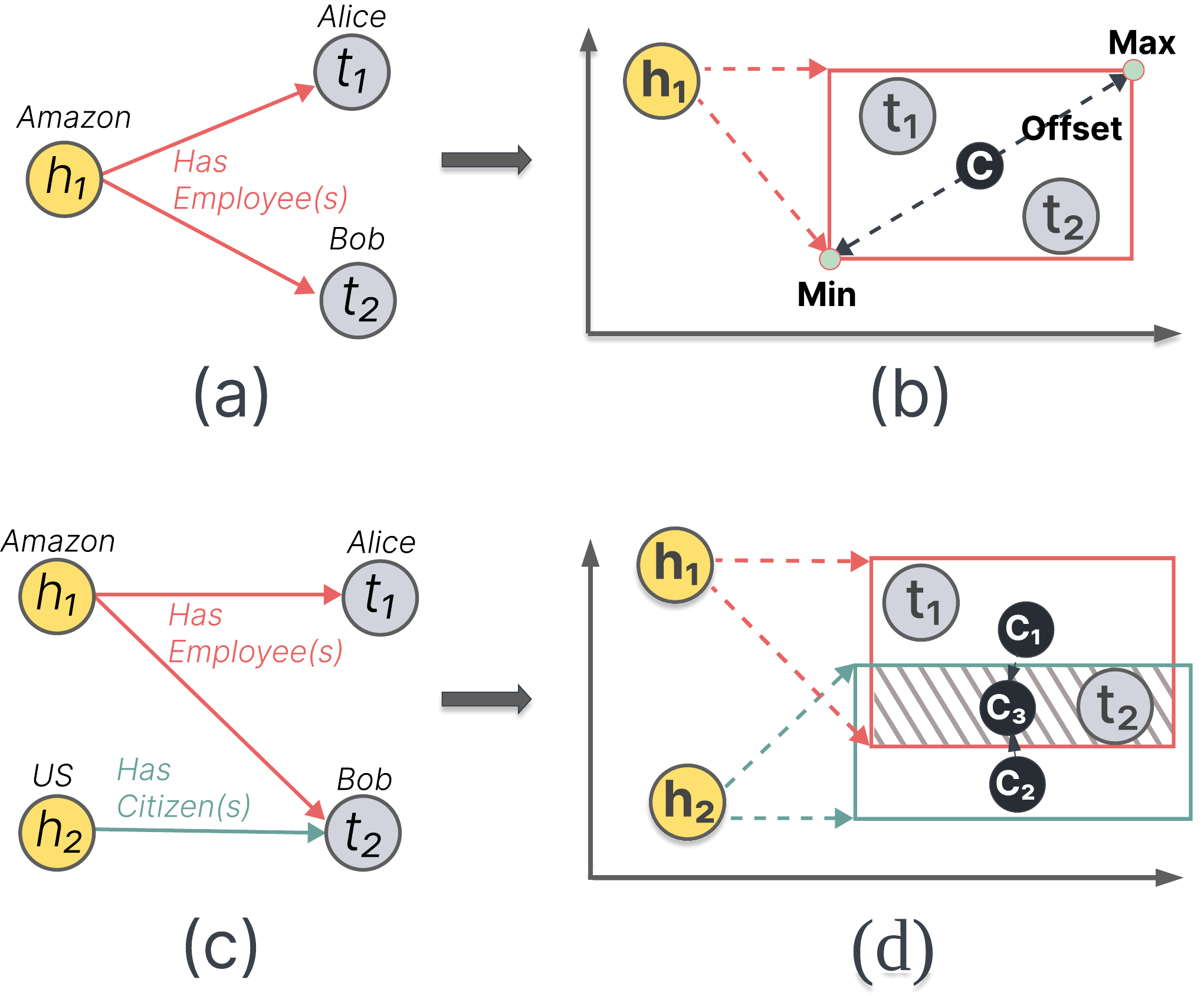}

    \caption{Illustration of the box embedding two operations: projection  and intersection in a knowledge graph (a, c) and vector space (b, d), respectively. $h_1$ and $h_2$ are the head entities representing a company Amazon and a country US, respectively. Similarly, $t_1$ and $t_2$ are tail entities representing two individuals in a KG. The red and green arrows are different relations between head entities and tail entities. Additionally, \textbf{h} and \textbf{t} are entities' vector forms in the vector space. Besides, \protect\component{\textbf{C}} denotes the center of the box. \textbf{Offset} represents the box range, thereby determining the box's size. \textbf{Max} and \textbf{Min} are the endpoints of the box's diagonal. (a) displays a 1-to-N relation triple in which \textit{Amazon} $(h_1)$  has \textit{employees} named \textit{Alice} $(t_1)$ and \textit{Bob} $(t_2)$. (b) demonstrates the 1-to-N triple computation in the 2-D vector space by the projection:  
    the relation \textit{(Has Employee(s))} projects the head entity into a box that contains the tail entities. 
    (c) indicates the intersection of two 1-to-N triples where \textit{Bob} $(t_2)$ is an \textit{employee} of \textit{Amazon} $(h_1)$ and a \textit{citizen} of \textit{US} $(h_2)$. (d) shows the box intersection in the 2-D vector space: two boxes' intersection gives rise to a smaller box that contains a tail entity (\textbfsub{t}{2}) relevant to both two triples. Otherwise, the irrelevant tail entity (\textbfsub{t}{1}) is outside the smaller box. The center \protect\component{\textbf{C}$_3$} of the smaller box is based upon the two projected box centers \protect\component{\textbf{C}$_1$} and \protect\component{\textbf{C}$_2$}.}
    \label{fig: operation}
    \vspace{-2em}
\end{figure}

\section{Background: Box Embedding in Knowledge Graphs}\label{sec:background}

\textbf{Knowledge Graphs and Relations.}
A knowledge graph models human knowledge as a directed graph with entities and relations. 
Each entity is a graph node, and each relation is a graph edge.
The relationship between any two entities is delineated by a triple $(h,r,t)$, where $h$ represents a head entity, $t$ represents a tail entity, and $r$ represents a relation.
Most relationships in a knowledge graph are 1-to-1 mappings, with the relation $r$ between the head entity $h$ and the tail entity $t$ being unique.
For example,
\textit{``US has a citizen named Bob''} is an example of the 1-to-1 relation, as shown in Figure~\ref{fig: operation}(c), and the corresponding triple is \textit{(US, Has Citizen(s), Bob)}. Besides, there are 1-to-N relations in a knowledge graph. 
For instance, \textit{``Amazon has employees Alice and Bob''} represents a 1-to-N relation. Figure~\ref{fig: operation}(a, c)  delineates the 1-to-N relation, where there are more than one tail entities
(i.e., \textit{Alice} and \textit{Bob}) for the same head entity and relation.
The triple is \textit{(Amazon, Has Employee(s), \{Alice, Bob\})}.

\textbf{Box Embedding.}
To facilitate computational manipulations on knowledge graphs, it's often necessary to apply Knowledge Graph Embedding (KGE) to represent entities and relations in KGs as continuous embedding vectors~\cite{wang2017knowledge}.
For the 1-to-1 relations in knowledge graphs, there have been many KGE methods to model them, e.g., TransE~\cite{bordes2013translating} and PTransE~\cite{lin2015modeling}.
However, they cannot model the 1-to-N relations that entail a set of tail entities as shown in Figure~\ref{fig: operation}(a).
Also, these methods cannot be used to define the intersection of multiple 1-to-N triples. The intersection of 1-to-N triples will obtain a set of common tail entities.
These common tail entities are relevant to all these 1-to-N triples.
For example, Figure~\ref{fig: operation}(c) shows the intersection of two triples, where \textit{``Bob is an employee of Amazon and also a US citizen"}.
To handle these challenges in a scalable manner, box embedding
 is introduced recently~\cite{vilnis2018probabilistic}.
Rather than representing a point in vector space, it represents an area and can handle 1-to-N relations and the intersection of 1-to-N relations using two operations: \textit{projection} and \textit{intersection}.
The projection operation of box embedding maps an entity embedding (i.e., a point) to a box area (i.e., axis-aligned hyper-rectangles) in the vector space (Figure~\ref{fig: operation}(b)). Tail entities should be enclosed within the projected box and satisfy the following condition:

\begin{equation}\label{formula: inbox}
    Box \equiv \{\text{v}\in\mathbb{R}^d: Cen(Box)-Off(Box) \preceq \textbf{\text{v}} \preceq Cen(Box)+Off(Box)\},
\end{equation}

where $\preceq$ denotes element-wise inequality, $Cen(Box)\in \mathbb{R}^{d}$ denotes the center point of the box and $Off(Box)\in \mathbb{R}^{d} \ge 0$, which stands for the positive offset of the box. The offset indicates the size of the projected box, as shown in Figure~\ref{fig: operation}(b).

The intersection operation of box embeddings models the intersection of multiple 1-to-N relations by intersecting several projected boxes.
For instance, Figure~\ref{fig: operation}(c) depicts the intersection of two triples, where 
$h_1$ and $h_2$ have different relations to $t_2$.
The intersection of two box embeddings projected from \textbfsub{h}{1} and \textbfsub{h}{2}, shown by the small shadowed box in Figure~\ref{fig: operation}(d), identifies the $t_2$ to which both $h_1$ and $h_2$ have relations.
\textbfsub{t}{2} is within the intersected box, indicating that  $t_2$ is the tail entities of both  two triples.

For visualization recommendations, it is common for a dataset to be represented as multiple visualizations types, which is a 1-to-N relation. 
To model these relations accurately, we utilize box embedding in our approach for adaptive visualization recommendations.

\section{Related Work}
The related work of this paper
can be categorized into three
groups:
visualization recommendation, knowledge graph embedding,
knowledge graph-based explainable recommendation.

\subsection{Visualization Recommendation}

Visualization recommendation aims to suggest or generate appropriate visualizations for a given dataset automatically and generally includes two types of methods~\cite{zhu2020survey}:
Rule-based methods and machine learning (ML)-based approaches.

Rule-based methods leverage visualization rules specified by visualization experts to recommend appropriate visualizations~\cite{mackinlay2007show,vartak2014seedb,wongsuphasawat2015voyager}.
For example,
Mackinlay~\textit{et al.} proposed \textit{Show Me}, which can automatically suggest visualizations using predefined visualization guidelines~\cite{mackinlay2007show}.
Using a predefined set of rules, voyager~\cite{wongsuphasawat2015voyager} and voyager2~\cite{wongsuphasawat2017voyager} enumerates all potential data columns in a dataset to get potential visualizations and further ranks all these visualizations to recommend appropriate choices. Additionally, Foresight~\cite{Demiralp2017ForesightRV} detects pre-defined statistical features from the dataset and then made recommendations according to these features.
and present them visually through appropriate chart types.
Though rule-based methods have been extensively studied, developing a comprehensive list of rules for visualization recommendations is challenging, and the maintenance of such empirical rules is often labour-intensive~\cite{li2021kg4vis}.

ML-based methods
learn
the mappings between input datasets and visualizations~\cite{zhu2020survey} from training examples.
For instance,
Vizdeck~\cite{key2012vizdeck} trains a linear model for this mapping.
DeepEye~\cite{luo2018deepeye} uses a \textit{learning-to-rank}~\cite{burges2005learning} model to rank visualization recommendations, then recommends the top scoring one.
Draco~\cite{moritz2018formalizing} employs the statistical model RankSVM to rank possible visualizations. 
More recently, deep neural networks have also been widely used for visualization recommendations, such as 
VizML~\cite{hu2019vizml}, Data2Vis~\cite{dibia2019data2vis} and Table2Charts~\cite{zhou2020table2charts}. 
While these ML-based methods can reduce the manual efforts of compiling rules for visualization recommendations, they often operate like a black box, making it difficult for general users to interpret them~\cite{zhu2020survey}.
Li~\textit{et al.~\cite{li2021kg4vis}} proposed a knowledge graph-based recommendation approach. Their approach recommends suitable visualization choices in a data-driven and explainable manner, making it the most relevant study to our work.
However, this approach fails to consider relationships between data columns in the dataset, which is crucial for determining visualization choices.

Unlike the above studies, \tool{} takes into account the cross-column relationships of the input dataset and can provide adaptive visualization recommendations and explanations.

\subsection{Knowledge Graph Embedding}\label{related work: embedding}

Knowledge Graph (KG) models the relations between different entities~\cite{ji2021survey},
and knowledge graph embedding (KGE) maps entities and relations into embedding vectors while preserving their semantic meanings, which mainly includes semantic matching models and translational distance models~\cite{wang2017knowledge}.
Semantic matching models evaluate the plausibility of a triple by matching the entities and relations with latent semantics in the vector space.
For example, RESCAL~\cite{nickel2011three} assigns a vector embedding to each entity in the knowledge graph, and each relation is interpreted as a matrix that models the semantic interaction between two entities. 
Dismult~\cite{yang2014embedding} restricts the relation matrix of RESCAL to multiple diagonal matrices, thereby simplifying the calculation of RESCAL.

Translational distance models use translation operations to represent the relations between any two entities, where the distance between the entity embedding after a translation and the other entity embedding indicates the plausibility of a triple.
TransE~\cite{bordes2013translating} is one of the most representative translational distance methods. 
When using TransE, combining the embedding vector of a head entity with that of a relation creates a new embedding in the vector space that approximates the tail entity.
The limitation of TransE is that it implicitly assumes that there are only one-to-one relationships between entities and cannot deal with 1-to-N relationships~\cite{wang2014knowledge}. 
Therefore, other methods have been proposed to improve the modeling of 1-to-N relationships in knowledge graphs~\cite{wang2014knowledge,chen2020knowledge,ji2015knowledge,ren2020query2box}.
For example, query2box~\cite{ren2020query2box} introduces box embeddings whereby a head entity embedding can be translated into a box by a relation embedding.
rather than a point in the vector space.
When the embedding vector of a tail entity lies inside the projected box embedding of a head entity, the corresponding triple is considered valid, making it able to represent 1-to-N relations.

Our approach is inspired by query2box~\cite{ren2020query2box} and incorporates box embedding in our knowledge graph to model the 1-to-N and intersection of 1-to-N relations in visualization recommendation. Also, we augment the original loss function of query2box to enhance the adaptability of recommended visualizations.

\subsection{Knowledge Graph Based Explainable Recommendation}

Knowledge graphs have been integrated into recommendation systems to enhance their interpretability~\cite{ji2021survey}.
According to the survey by Li~\textit{et al.}~\cite{li2020survey}, the knowledge graph-based explainable recommendation methods can be grouped into two categories:
internal route-based methods and external route-based methods.
For the internal route-based methods, the recommendation algorithms are designed by explicitly considering
the knowledge graphs, including their entities, relations, paths, and rules, to improve the recommendation performance and provide explanations.
For instance,
Wang~\textit{et al.}~\cite{wang2019explainable} defined the relations between entities as sequential paths and further leveraged a Recurrent Neural Network (RNN) to model the sequential dependencies of entities within a knowledge graph.
Also, Ma~\textit{et al.}~\cite{ma2019jointly} directly derived recommendation rules from the knowledge graph and recommended items based on the extracted rules.

In contrast, external route-based recommendation methods are not built upon knowledge graphs. Instead, they only use external knowledge graphs to generate explanations for the recommendation results.
For example, the medical knowledge graph has been used to discover possible explanations for previous medical treatments~\cite{seneviratne2019enabling}. 
Also, Sarker~\textit{et al.}~\cite{sarker2017explaining} utilized an external knowledge graph to elucidate the behaviors of neural network classifications.

Our approach falls under internal routes-based recommendation methods, and integrates a knowledge graph into our recommendation framework.
Tracing back paths in the knowledge graph can provide meaningful explanations for the recommended visualizations.

%% file: src/3-method.tex
\begin{figure}[!ht]
    \centering
    \includegraphics[width=\linewidth]{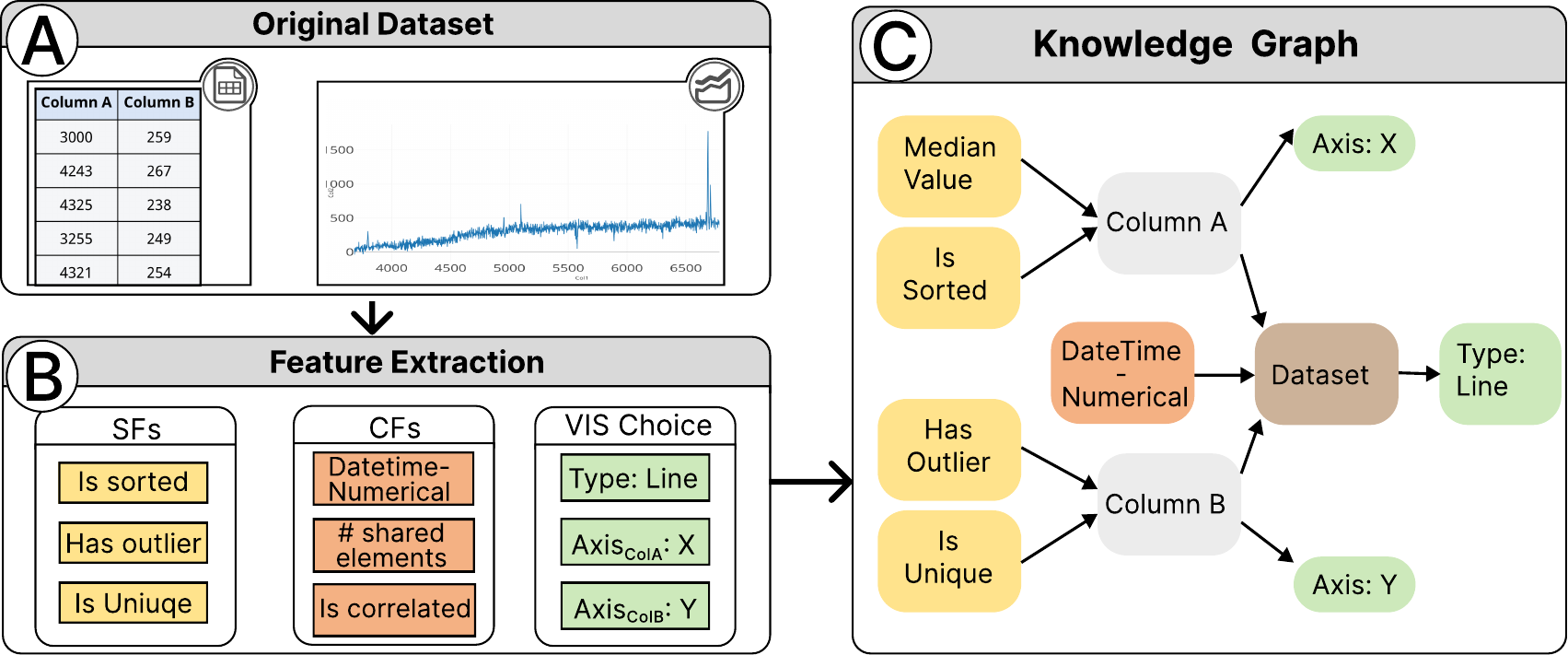}
    \caption{An example of transforming a dataset into a knowledge graph. \protect\subcomponent{\textbf{A}} The dataset contains a pair of tabular data and corresponding visualization. \protect\subcomponent{\textbf{B}} With feature extraction, the individual data columns' characteristics, namely single-column features (SFs) and the interrelationships of two data columns, namely cross-column features (CFs) will be obtained, as well as the mapping between the dataset and visualization choices (VIS Choice). Visualization choices include the visualization type (e.g., line chart) and the axis (e.g., x-axis). \protect\subcomponent{\textbf{C}} The \SF{single-column features}, \CF{cross-column features}, \COL{data columns}, \DS{datasets}, and \VIS{visualization choices} are represented as entities in the knowledge graph.
    Only part of the features and entities are shown.}
    \label{fig: kg_construction}
    \vspace{-2em}
\end{figure}

\section{Our Method}\label{sec:tech}

We propose \tool{}, an adaptive and explainable knowledge-graph-based approach to recommend visualizations for tabular datasets.
Given that the choice of standard visualization types (i.e., line chart, bar chart, box plot, and scatter plot) often depends on the
two data columns displayed on the chart axes, we formulate the visualization recommendation problem for two-dimensional datasets as logical reasoning over the KG to infer
visualization types for two-column datasets\cite{ren2020query2box,hamilton2018embedding}.
The source code for our approach is \href{https://github.com/AlexanderZsh/AdaVis}available.

\begin{figure}[!ht]
    \centering
    \includegraphics[width=\linewidth]{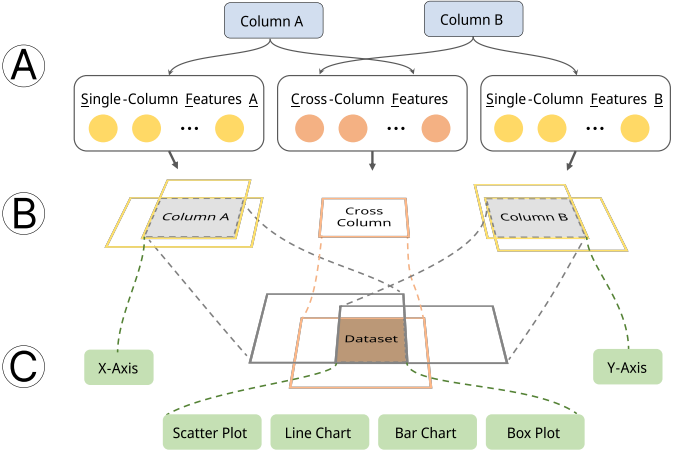}
    \caption{ This figure illustrates the workflow of \tool{} infers appropriate visualization types. \protect\subcomponent{\textbf{A}} illustrates that we extract single-column and cross-column features from a dataset. Each node represents a feature entity.
    \protect\subcomponent{\textbf{B}} describes the procedure for generating data column and dataset box embeddings in inference. A hollow yellow/orange rectangle represents a box embedding from a single-column/cross-column feature. A gray rectangle denotes a data column obtained from an intersection of box embeddings of single-column features. The brown rectangle represents the dataset generated from an intersection of box embeddings that represent data columns and cross-column features.
    \protect\subcomponent{\textbf{C}} indicates that box embeddings of data columns and dataset can infer correct visualization choices (i.e., axis and visualization types).}
    \label{fig: overview}
    \vspace{-1.8em}
\end{figure}

\subsection{Overview}\label{sec:method_overview}
When determining appropriate  visualizations for a dataset, users often need to consider the characteristics of two data columns of interest and their interrelationships. 
The design of \tool{} is inspired by the logical reasoning process of humans when they select the right visualizations. Such a reasoning process is modeled by a knowledge graph consisting of entities (i.e., single-column features,  cross-column features, data columns, datasets, and visualization choices). In this paper, we refer to an individual column of a dataset as a data column. A single-column feature is a quantified characteristic of a data column. Similarly, a cross-column feature is a quantified interrelationship between data columns.

\tool{} comprises feature extraction, knowledge graph construction, box embedding learning, model inference,
and explanation generation for the visualization recommendation, which collectively facilitate visualization recommendation.
Given a corpus of dataset-visualization pairs, \tool{} can learn the implicit mapping between datasets and visualization choices (i.e., visualization types and an axis), which will be further used for visualization recommendations for new datasets.
Specifically, as shown in Figure~\ref{fig: kg_construction} \protect\subcomponent{\textbf{A}} and \protect\subcomponent{\textbf{B}}, \tool{} extracts related features for columns in a dataset.
Data column features, data columns, datasets, and visualization choices will be used to construct knowledge graphs (Figure~\ref{fig: kg_construction} \protect\subcomponent{\textbf{C}}).
Further, we utilize the box embedding technique~\cite{ren2020query2box} to learn the embeddings of entities in the knowledge graph, which can well model their relations in visualization recommendations.
We will use the learned embeddings of the knowledge graph's entities and relations to infer suitable visualization choices (i.e., visualization types or an axis) for an unseen dataset. Moreover, a template-based explanation module, built upon the knowledge graph, is integrated into \tool{} to generate natural language explanations for the visualization recommendation.

\subsection{Feature Extraction}\label{sec:feature extraction}

Visualization choices depend on the characteristics of the input dataset.
To extract quantified characteristics (features) of datasets, we first surveyed prior studies of visualization recommendation~\cite{key2012vizdeck,luo2018deepeye,hu2019vizml} and visualization insight discovery~\cite{ding2019quickinsights,harris2021insight}.
Based on the survey results, we categorized the dataset features
into two types: \textit{single-column features} and \textit{cross-column features}, as shown in Figure~\ref{fig: kg_construction}  \subcomponent{\textbf{A}} and \subcomponent{\textbf{B}}. 

The \textit{single-column features} describe the properties of individual data columns, such as the column length, the mean and variance of a column's values.
We
extracted 80 distinct single-column features 
that can collectively model
the properties of individual data columns from various perspectives. 
Besides the 80 single-column features, we also extracted 40 well-designed cross-column features in \tool{} by referring to prior studies~\cite{luo2018deepeye,hu2019vizml,ding2019quickinsights}.
The \textit{cross-column features} capture the interrelationships between two columns, e.g., two columns' data types are categorical and numerical.
Such cross-column features are also crucial for deciding the visualization choices. For example, compared with a line chart, it is more appropriate to use a scatter plot to visualize a two-column dataset with two columns exhibiting significant correlation~\cite{cui2019datasite}.
 Furthermore, as shown in Figure~\ref{fig: kg_construction} \protect\subcomponent{\textbf{B}}, we obtain the visualization choices (i.e., visualization types and axes) of datasets.
A complete list of all the features used in \tool{} can be found in the appendix.

\begin{table*}[hbt!]
    \centering
    \small
    \caption{The table shows two categories of triples in the knowledge graph: 1-to-N and the intersection of 1-to-N. 1-to-N triples can be classified as five relations in the knowledge graph. The intersection of 1-to-N triples intrinsically combines multiple 1-to-N relations. 
    }

    \setlength{\aboverulesep}{0.5pt}
    \setlength{\belowrulesep}{0.5pt}
    \begin{tabular}{p{1.5cm}|p{3.3cm}|p{4.8cm}|p{6.5cm}}
        \toprule
        Type & Relations & Meanings & Examples\\
        \midrule
        \multirow{5}{*}{1-to-N}&
        $\mathbb{R}_{SF\rightarrow COL}$ & Data columns with the specific single-column feature are & The column (\mbox{\tiny$COL$}) length is 50 (\mbox{\tiny${SF}$})\\\cmidrule{2-4}
        &$\mathbb{R}_{CF\rightarrow DS}$ & Datasets with the specific cross-column feature are & Percentage of unique values (\mbox{\tiny${CF}$}) shared in a Dataset 's two columns (\mbox{\tiny${DS}$}) \\\cmidrule{2-4}
        &$\mathbb{R}_{COL\rightarrow DS}$ & Data columns in the specific dataset are & A column representing grades (\mbox{\tiny${COL}$}) in a dataset (\mbox{\tiny${DS}$}) about students' grades (Figure~\ref{fig: gpa}\subcomponent{\textbf{A}}), \\\cmidrule{2-4}
        &$\mathbb{R}_{COL\rightarrow VIS_{Axis}}$ & The data columns can be encoded on & A column represents grade (\mbox{\tiny${COL}$}) is encoded on the y-axis  (\mbox{\tiny${VIS_{Axis}}$}) in a visualization (Figure~\ref{fig: gpa}\subcomponent{\textbf{C}})\\\cmidrule{2-4}
        &$\mathbb{R}_{DS\rightarrow VIS_{Type}}$ & Visualization types available for the dataset are  & The dataset about students' grades (\mbox{\tiny${DS}$}) is visualized as a box plot (\mbox{\tiny${VIS_{Type}}$}) (Figure~\ref{fig: gpa}\subcomponent{\textbf{C}})\\
        \midrule
        \multirow{2}{*}{\shortstack{Intersection \\ of \\ 1-to-Ns}}& $\mathbb{R}_{SF_1\rightarrow COL} \cap \mathbb{R}_{SF_2\rightarrow COL} \cap \dots \cap \mathbb{R}_{SF_n\rightarrow COL}$ & Data columns with $n$ single-column features are & A set of columns  (\mbox{\tiny${COL_1}\dots{COL_n}$})  have the same features such as column length is 50  (\mbox{\tiny${SF_1}$}), column values are sorted (\mbox{\tiny${SF_2}$}) and so on (\mbox{\tiny${SF_3}\dots{SF_n}$})\\\cmidrule{2-4}
        &$\mathbb{R}_{COL_1\rightarrow DS} \cap \mathbb{R}_{COL_2\rightarrow DS} \cap  \mathbb{R}_{CF_1\rightarrow DS} \cap \dots \cap \mathbb{R}_{CF_m\rightarrow DS}$ & Datasets including two specific data columns and a set of $m$ cross-columns features are & A set of datasets (\mbox{\tiny${DS_1}\dots{DS_n}$}) have the same characteristics: their columns' characteristics (\mbox{\tiny$({COL_{11}}, {COL_{12}})\dots({COL_{n1}}, {COL_{n2}})$}) are the same, and these datasets have the same cross-column features such as the pairwise columns have overlapping value ranges (\mbox{\tiny$CF_1$}) and so on (\mbox{\tiny$CF_2\dots CF_m$})\\
        \bottomrule       
    \end{tabular}
    \label{table: triples}
    \vspace{-2em}
\end{table*}

\subsection{Knowledge Graph Construction}\label{sec: method_KGconstruction}

A knowledge graph allows us to model the mapping between datasets and different visualization choices.
With a well-designed knowledge graph, we can further recommend appropriate visualizations.

\textbf{Definition of Entities.} As shown in
Figure~\ref{fig: kg_construction} \protect\subcomponent{\textbf{C}},
we define five classes of entities that are encoded with different colors: single-column features~($\mathbb{E}_{SF}$, yellow nodes), data columns~($\mathbb{E}_{COL}$, gray nodes), datasets~($\mathbb{E}_{DS}$, brown nodes), cross-column features~($\mathbb{E}_{CF}$, orange nodes) and visualization choices~($\mathbb{E}_{VIS}$, green nodes).

As shown in Table~\ref{table: triples}, $\mathbb{E}_{SF}$ represents features extracted from individual data columns; 
$\mathbb{E}_{CF}$ are cross-column features; 
$\mathbb{E}_{COL}$ and $\mathbb{E}_{DS}$ refer to data columns and datasets, respectively; $\mathbb{E}_{VIS}$ refers to the choices available for visualizations, and consists of four popular charts (i.e., bar chart, line chart, scatter plot, and a box plot)~\cite{battle2018beagle}. as well as the two commonly used axes (i.e., the x-axis and y-axis).

Since single-column and cross-column features can be continuous values, we discretize them into different intervals to represent them as entities in a knowledge graph.
Specifically, we utilize the widely-used MDLP approach~\cite{liu2002discretization} to transform continuous features into categorical features.

\textbf{Definition of Relations.} 
As illustrated in Table~\ref{table: triples}, there are five relations classes in our knowledge graph.
(1) $\mathbb{R}_{SF\rightarrow COL}$ denotes a class of relations that associate single-column features with single data columns, and this class indicates that these features are present in a single data column. (2) $\mathbb{R}_{COL\rightarrow DS}$ represents a class of relations that link a single data column
to a dataset. It shows that datasets contain the data column. (3) Similar to $\mathbb{R}_{SF\rightarrow COL}$, $\mathbb{R}_{CF\rightarrow DS}$ is a class of relations that indicate cross-column features exist in datasets. For example, $\mathbb{R}_{CF\rightarrow DS}$ means that \textit{``(one cross-column feature) exists in columns of (a dataset)''}. (4) $\mathbb{R}_{COL\rightarrow VIS}$ shows that single data columns are encoded with a specific axis. For example, $\mathbb{R}_{COL\rightarrow VIS}$ means that \textit{``(one data column) is encoded as (x-axis)''}. (5) Similarly, $\mathbb{R}_{DS\rightarrow VIS}$ means a dataset is encoded as a visualization type.

\textbf{Definition of Triples.} 
After defining entities and relations, we generate triples based on existing dataset-visualization pairs. These triples are instances of the defined knowledge graph relations. These triples can be categorized into two types, 1-to-N and intersection of 1-to-Ns, as illustrated in Table~\ref{table: triples}. As shown in Figure~\ref{fig: kg_construction} \subcomponent{\textbf{C}}, a 1-to-N triple is constructed by a relation (an arrow) and two types of entities (two nodes with different colors). It is a 1-to-N triple (N$\ge 1$) because one head entity may correspond to multiple tail entities by a relation. For example, in Figure~\ref{fig: kg_construction} \subcomponent{\textbf{C}}, ``(Has Outlier $\xrightarrow{}$ Column B)'' denotes a single-column feature (i.e., Has Outlier) could exist in many data columns (i.e., Column B), so, in a triple, the single-column feature (i.e., head) may correspond to many data columns (i.e., tails) by a relation (i.e., $\mathbb{R}_{SF\rightarrow COL}$). 
There are five types of 1-to-N triples since the knowledge graph contains five types of relation. 
 Besides 1-to-N triples, an intersection of 1-to-N triples is also generated from the knowledge graph. For example, in Figure~\ref{fig: kg_construction} \subcomponent{\textbf{C}}, a combination of two triples, i.e.,``(Has Outlier $\xrightarrow{}$ Column B) \& (Is unique $\xrightarrow{}$ Column B)'', is an instance of intersection of 1-to-N triples. The example refers to a set of data columns with both features (i.e., Has Outlier \& Is unique). 


\subsection{Box Embedding Learning}

Box Embedding Learning guides \tool{} to learn possible visualization choices for a dataset.
As introduced in Section~\ref{sec:background}, a head entity and a relation in triples are projected onto a box embedding. For example, suppose a single-column feature exists in data columns. In that case, this condition can be regarded as a triple like \textit{``(A single-column feature, Exists in, Some data columns)''}. As illustrated in Figure~\ref{fig: overview} \subcomponent{\textbf{A}} and \textbf{\subcomponent{\textbf{B}}}, the single-column feature (a yellow node) is
projected into a
box embedding (a hollow yellow rectangle) and its tail entities (i.e., data columns have single-column features) are supposed to lie within the box embedding. Besides transforming triples into box embeddings, multiple boxes from multiple triples can be merged to create a smaller box embedding. In Figure~\ref{fig: overview} \subcomponent{\textbf{B}} and \textbf{\subcomponent{\textbf{C}}}, for instance, the box embeddings of columns (gray rectangles) and a cross-columns feature (a hollow orange rectangle) are intersected to obtain a smaller box embedding that represents datasets. The dataset entities with those columns and the cross-column feature will be inside the dataset box embedding. The distance between the box and tail entity is defined as:

\begin{equation}\label{formula:dist_sum}
        \text{dist}_\text{box}(t;b) = \text{dist}_\text{outside}(t;b) + \alpha \cdot\text{dist}_\text{inside}(t;b) + \beta\cdot b_\text{size},
\end{equation}

where $b$ denotes the box embedding from the head and relation in a triple, and $t$ represents an embedding of the tail entity. The $\text{dist}_\text{box}(t;b)$ serves as a scoring function that measures the distance in vector space between the tail entity's the embedding and the box embedding. 
The distance function can be decomposed into three sub-functions: $\text{dist}_\text{outside}$ identifies whether the tail entity $t$ is within the box $b$. If $t$ is inside the box, the score of $\text{dist}_\text{outside}$ is 0. Otherwise, $\text{dist}_\text{outside}$ returns the distance between the tail entity $t$ and the close side of the box $b$. 
As for $\text{dist}_\text{inside}$, it calculates the distance between the center of the box $b$ and $t$
(or the distance between the close side of the box and $t$ if $t$ is outside the box). 
The hyper-parameter $\alpha \in [0,1]$ controlls the weight of $\text{dist}_\text{inside}$. If $\alpha = 0$,  $\text{dist}_\text{inside}$ is nullified, causing the scoring function to solely consider the distance of the tails from the box $b$. Furthermore,, $b_{\text{size}}$ is designed to control the box size in case the box size is so large that it includes irrelevant tail entities. Thus, $\beta \in[0,1]$ is a hyparameter that controls the box size.
In summary, $\text{dist}_\text{box}(t;b)$ aims to measure how far a tail entity is from the box embedding.
$\text{dist}_\text{outside}$, $\text{dist}_\text{inside}$ and $b_{\text{size}}$ are defined as follows:

\begin{equation}\label{formula:dist_out}
        \text{dist}_\text{outside}(t;b) = ||\text{Max}(t-b_\text{max},0) + \text{Max}(b_\text{min}-t,0)||_1
\end{equation}
\vspace{-1mm}
\begin{equation}\label{formula:dist_cen}
        \text{dist}_\text{inside}(t;b) = ||\text{Cen}(b) - \text{Min}(b_\text{max}, \text{Max}(b_\text{min},t))||_1,
\end{equation}


\begin{equation}\label{formula: box}
    \begin{aligned}
        b_\text{size}(b) & = ||b_\text{max} - b_\text{max}||_2,\\
    \end{aligned}
\end{equation}

where $b_\text{max}$ and $b_\text{min}$ represent endpoints of the box $b$, as delineated in Figure~\ref{fig: operation}(b), Cen($b$) denotes the box's center point. $b_\text{size}$ represents the box size.

In each iteration of model training, we sample a set of positive and negative triples from the training dataset. In the knowledge graph, positive triples are the correct triples, while negative triples are incorrect triples whose tail entities $t$ do not correspond to the head and relation. For example, \textit{(US, Has Citizen, UK)} is a negative sample where the tail entity (\textit{UK}) is not the answer to the head entity (\textit{US}) and relation (\textit{Has Citizen}). We generate $k$ negative samples for a positive triple. The positive and negative samples constitute a minibatch of training samples. In this minibatch, \tool{} is updated by the calculated loss. The loss function is defined as follows, according to~\cite{ren2020query2box}:

\begin{equation}\label{formula:loss}
    L = -\text{log}\sigma (\gamma-\text{dist}_\text{box}(b;t))-\sum^k_{i=1}\frac{1}{k}\text{log}\sigma(\text{dist}_\text{box}(b;t'_i)-\gamma),
\end{equation}

where $\sigma$ is the Sigmoid function, and $\gamma$ is a fixed scalar margin. $t$ means a positive tail entity, while $t'_i$ refers to a negative tail entity that should be far away from the box embedding $b$. The intuition behind the loss function is that the correct answer (positive tail) should lie inside the box and as close to the box center as possible, but the incorrect answer (negative tail) should be far away from the box as possible.

\subsection{Model Inference}\label{sec:model_inference}
In the training step, the model learned the embeddings of entities and relations in the knowledge graph. 
This section clarifies how \tool{} uses the learned embeddings to infer axes and possible  visualization types for an unseen dataset.

\tool{} extracts single-column and cross-column features from the unseen dataset, as illustrated in Section~\ref{sec:feature extraction}. 
Our knowledge graph contains entities of these features, and the corresponding embeddings of these features have been learned.
Then, these single-column features are projected to box embeddings, as shown in Figure~\ref{fig: overview} \subcomponent{\textbf{A}} and \subcomponent{\textbf{B}}.
\tool{} further obtains the box embedding of the data column from an intersection of single-column features' box embeddings.
Each single-column feature represents a particular characteristic of a data column, so the intersected box embedding of single-column features represents a data column's overall characteristics. 
Having obtained the data column embedding, we can infer an appropriate axis for the data column. In the paper, a dataset includes two columns, and its visualizations are also two-dimensional, with one x-axis and one y-axis.
In other words, one data column only corresponds to one axis. Due to this fact, we should determine which axis is best suited to this data column. To this end, we first infer the axis' box embedding ($Box_{Axis}$) from this data column's box embedding ($Box_{COL}$). $Box_{COL}$ represents the data column, and $Box_{Axis}$ represents the optimal axis to this data column. The axis embedding ($Box_{Axis}$) is obtained from the data column embedding ($Box_{COL}$) by a relation ($\mathbb{R}_{COL\rightarrow VIS_{Axis}}$) which specifies the transformation from a data column to its optimal axis. 
Since $Box_{Axis}$ represents the optimal axis for this column and two axis entities (i.e., x, y-axis) exists in the constructed knowledge graph, we can
calculate the distance between $Box_{Axis}$ and the embeddings of these two axis entities. The distance measures the plausibility between the optimal axis and axis entities, and the axis entities are denoted by $\mathbb{E}_{VIS_{Axis}}$ (Table~\ref{table: triples}). Additionally. the distance calculation is done by Equation~\ref{formula:dist_sum}. A lower calculated score means higher plausibility.
For example, if $\text{dist}_\text{box}(\mathbb{E}_{VIS_{x}};Box_{Axis})<\text{dist}_\text{box}(\mathbb{E}_{VIS_{y}};Box_{Axis})$,  \tool{} chooses x-axis for the data column because the data column's optimal axis is more plausible to the x-axis entity than the y-axis entity.

As for inferring visualization types to a dataset, we will identify a set of visualization types that are appropriate to the dataset. Unlike the 1-to-1 mapping between a data column and an axis, a 1-to-N mapping exists between a dataset and multiple visualization types, as multiple visualization types are suitable for the same dataset (Figure~\ref{fig: gpa}). To infer visualization types, we first need to obtain the dataset representation in terms of box embedding. In a manner similar to data column embedding obtainment, a box embedding of the dataset ($Box_{DS}$) is gained by intersecting the box embeddings of its data columns and cross-column features, as shown in Figure~\ref{fig: overview} \subcomponent{\textbf{B}} and \subcomponent{\textbf{C}}. From the $Box_{DS}$, we infer its optimal visualization types ($Box_{Type}$) in terms of box embedding by a relation $\mathbb{R}_{COL\rightarrow{} VIS_{type}}$.
Since one dataset may have more than one suitable visualization type, we need to classify these suitable visualization types at once. Due to this requirement, Equation~\ref{formula: inbox} is used to identify which visualization types entities (e.g., line, bar) are inside $Box_{Type}$. In other words, if visualization type entities are appropriate for the dataset, they will be enclosed by the $Box_{Type}$.

\begin{figure}[!ht]
    \centering
    \includegraphics[width=\linewidth]{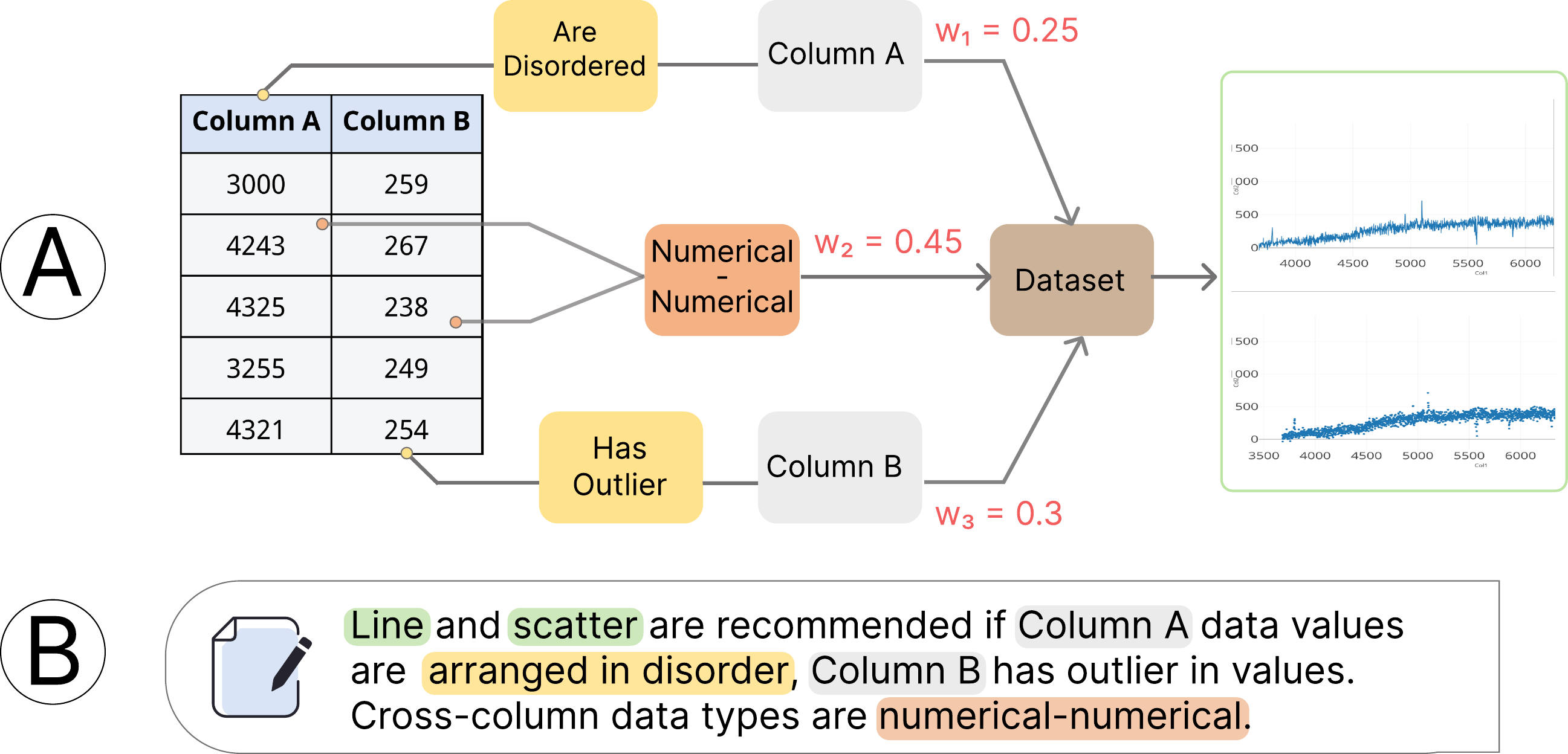}
    \caption{An illustration of explanation generation for a visualization type recommendation. \protect\subcomponent{\textbf{A}} a visualization of three paths with quantified importance for the recommendation result. Each path represents a single-column feature (yellow) or cross-column feature (orange). Their quantified importance in the recommendation result is denoted by w$_i$. The w$_i$ are obtained by normalizing their attention scores. Only parts of the paths are shown. \protect\subcomponent{\textbf{B}} an explanation for the recommendation. The features are filled into the explanation template's slot.}
        \label{fig: explanation}
\vspace{-2em}
\end{figure}

\subsection{Explanation Generation}\label{sec:explanation}
In this section, we will illustrate how \tool{} provides fine-grained explanations for the recommendation of a specific dataset. As mentioned in Section~\ref{sec:model_inference}, the procedure of \tool{} inference is sequential and follows an inference path, such as  $\text{\{Are disordered\}}\rightarrow
\text{\{Column A\}} \rightarrow\text{\{Dataset\}} \rightarrow \text{\{Line, Scatter\}}$ (Figure~\ref{fig: explanation} \protect\subcomponent{\textbf{A}}). The inference path indicates: 
\textit{\textbf{If} data values are \textbf{\{arranged in disorder\}} in \textbf{\{Column A\}}, \textbf{then} the \textbf{\{Dataset\}} can be visualized as \textbf{\{Line, Scatter\}}}. 
As an inference path from a single-column feature (e.g., the column values are not ordered)
or a cross-column feature (e.g., two data column's data types are both numerical)
to the dataset, this path contributes to the prediction of the visualization type.
To quantify each path's importance to the inference result,  \tool{} leverages the attention mechanism in the inference paths~\cite{bahdanau2014neural}. We input box embeddings of single or cross-column features into a fully-connected neural network (i.e., MLP) and obtain output values as feature attention scores. 
The path's importance to the inference result is represented by its attention score, as shown in Figure~\ref{fig: explanation} \protect\subcomponent{\textbf{A}}.
With the quantified importance, we can reversely trace the important inference paths and reach specific single-column and cross-column feature entities.

We have developed a rule-based template to automatically translate important features into natural language sentences. We designed an explanation template. The template is a pre-defined sentence with empty slots: `[\textit{VIS}] is recommended if [\textit{Column}] has [\textit{Single-column features}], and Cross-column (i.e., the relationship between two columns)
has [\textit{Cross-column features}]'. 
\tool{} recommends visualization results for the dataset and determines features that are critical to the final visualization recommendation. For example, Figure~\ref{fig: explanation} \protect\subcomponent{\textbf{B}} provides an example of the template instance where the crucial features (color-filled texts) replace the slots. Both the recommended visualization types and important features are filled in on the template. 
In particular, according to the principles advised by Yuan \textit{et al.}~\cite{yuan2022isea}, we propose detailed rules for our template to enhance the interpretability of the resulting explanation for humans.
(1) We limit an explanation’s maximum number of features to four to prevent cognitive overload. 
Also, to ensure the features in the explanation are not trivial to the recommendation result, we filter out unimportant features based on their importance scores which are calculated by the attention score. 
(2) We also empirically filter out some features which are statistically informative but could confuse users, such as Moment 9, Gini coefficient, and Kurtosis.
(3) We divide numerical features' semantics into several degrees based on their discretized intervals. For example, if a single-column feature such as \textit{column length} is discretized into two intervals by MDLP, and \textit{``a data column length is equal to 5''} is within the lower interval, the column's length will be regarded as short. 
(4) We avoid including categorical features that have a negative value as they are not relatable to an understandable concept for users.
For instance, \textit{``There is no linear regression"} is a negative feature value and does not make sense to users.

%% file: src/4-evaluation.tex
\section{Evaluation}
To demonstrate the effectiveness of \tool{}, we extensively evaluated \tool{} with quantitative comparisons, case studies, and expert interviews. 
This section  introduces 
the setup of our experiments (Section~\ref{sec:experiment setup}) and the results (Sections~\ref{sec:quantitative} and \ref{sec:qualitative}) in detail.

\subsection{Experiment Setup}\label{sec:experiment setup}

This section introduces the dataset used for evaluation and the model settings in our experiments.

\textbf{Corpus.} 
We have used the large-scale corpus of visualization-dataset pairs introduced in VizML~\cite{hu2019vizml} to evaluate the effectiveness of \tool{}.
The VizML corpus is crawled from Plotly Chart Studio~\footnote{\url{https://plot.ly/online-chartmaker/}}.
In the corpus, each visualization-dataset pair contains one dataset and the corresponding visualization created by users.

We first filtered all visualization-dataset pairs with the dataset consisting of two data columns from the corpus.
Then, we retained four types of visualizations, i.e., bar charts, scatter plots, line charts, and box plots, since they are commonly used in Plotly~\cite{battle2018beagle}
and are standard chart types.
Our final dataset consists of approximately 30000 dataset-visualization pairs, which are further randomly divided into training and testing sets by a ratio of 2:1.

\subsection{Quantitative Evaluation}\label{sec:quantitative}
We conducted experiments to evaluate \tool{} quantitatively from three perspectives: single-class visualization recommendation (i.e., an axis and a visualization type), multiple-class visualization recommendation (i.e., multiple visualization types), and the validity of cross-column features.

\subsubsection{Single-class Visualization Recommendation}
\label{sec-single-class-vis-recommendation}

We conducted an experiment to assess our method. As mentioned in Section~\ref{sec:method_overview}, a visualization consists of axes and a visualization type, so we evaluate \tool{} by carrying out two tasks: (1) 
axis recommendation
for a data column; (2) 
visualization type recommendation
for a dataset. Our corpus was collected from Plotly Chart Studio, where users typically create one visualization for each dataset. Therefore,
the visualization choice is single-class in the experiment. To be specific, our experiment took one axis and one visualization type as the ground truth for each data column and dataset, respectively.

\textbf{Baseline Models.}
We compare \tool{} with three baseline models: KG4Vis~\cite{li2021kg4vis}, GQE~\cite{hamilton2018embedding} and  Decision Tree. Among them, KG4Vis~\cite{li2021kg4vis} is the most relevant to our approach, as it also employs a knowledge graph for recommending visualization choices
and further providing explanations for its recommendations.
The scores attributed by KG4Vis to each visualization choice were derived by taking an average of the inference results across all columns of a dataset. 
Besides KG4Vis, we also compare \tool{} to two other models: GQE~\cite{hamilton2018embedding} and the Decision Tree approach. GQE is a widely used model in knowledge-graph recommendation-based tasks~\cite{ren2020beta, zhang2021cone, hu2022type}, which makes it a suitable baseline to be compared with our approach.
The Decision Tree model is essentially an explainable ML model and has also been employed in visualization recommendation~\cite{hu2019vizml, luo2018deepeye}.
The models in our experiment assign a score to each visualization option and rank them in descending order. For this, \tool{} employs Equation~\ref{formula:dist_sum} to calculate the score.

\textbf{Metrics.}  We applied two widely used metrics to evaluate the performance of our method comprehensively: Mean Rank (MR) and Hit@2~\cite{bordes2013translating}. 
MR represents the average ranking of the correct visualization choices (the lower the MR score, the better performance), and Hits@2 represents the proportion of correct visualization choices that rank in the top two inference results (the higher the Hits@2 score,  the better performance). Since the axis is either the x-axis or the y-axis, we use accuracy to evaluate its binary prediction (the higher the accuracy score, the better performance).

\textbf{Result and Analysis.}
Table~\ref{table:quantitative_evaluation} shows that
\tool{} outperforms the baseline models in recommending appropriate visualization types, underscoring the effectiveness of 
\tool{}.
A contributing factor to \tool{}'s performance is its use of cross-column features and the intersection of box embeddings which can effectively model the intersection of all features, 
which extracts critical insights from both single-column and cross-column features. 
These insights are subsequently employed to recommend suitable visualization types.
As for the axis prediction, Decision Tree marginally surpasses \tool{}. However, it is important to note that retraining Decision Tre\alexin{e} for different tasks needs an additional computational burden. In contrast, \tool{} is trained only once for both tasks.

\begin{table}[ht!]
\caption{The table displays the quantitative result regarding visualization type and axis recommendation among \tool{} and baseline models.}
\centering
\vspace{1em}
\setlength{\aboverulesep}{0.5pt}
\setlength{\belowrulesep}{0.5pt}
\begin{tabular}{llll}
\toprule
 &   Axis & \multicolumn{2}{l}{Visualization Types} \\ \cmidrule{2-4} 
 &   Accuracy & MR & Hits@2 \\ \midrule
 \multirow{1}{*}{\tool{}} 
&  0.8536  & \textbf{1.626} & \textbf{0.8421} \\ \midrule
\multirow{1}{*}{GQE} 
& 0.8487 & 1.884 & 0.7268  \\ \midrule

 \multirow{1}{*}{KG4Vis} &  0.7579  & 1.736 & 0.8111 \\
\midrule
\multirow{1}{*}{Decision Tree} & \textbf{0.8795} & 1.893 & 0.7189 \\ 
\bottomrule
\end{tabular}
\label{table:quantitative_evaluation}
\vspace{-1em}
\end{table}

\begin{table}[h]
\setlength{\aboverulesep}{0.5pt}
\setlength{\belowrulesep}{0.5pt}
\caption{The table displays \tool{} recommendation effectiveness for multiple answers.}
\resizebox{\linewidth}{!}{
\begin{tabular}{lcccccc}
\toprule
               & \multicolumn{6}{c}{Visualization Types (Adaptability)}                         \\ \cmidrule{2-7} 
               & \multicolumn{3}{c}{Two Choices}               & \multicolumn{3}{c}{Three Choices} \\ \midrule
               & Recall & Precision & \multicolumn{1}{c|}{F1} & Recall   & Precision  & F1      \\ \midrule
\tool{}   &  \textbf{0.6084} &  0.8259    &  \textbf{0.6621}                  &  \textbf{0.6147}  &  0.7943     &  \textbf{0.6596}  \\ \midrule
 GQE &  0.4002 &  0.8005   &  0.5337                  &  0.2648  &  0.7743    &  0.3972  \\ \midrule
KG4Vis &  0.304 &  0.608   &  0.406  &  0.3333  &  \textbf{1.0}  &  0.5  \\ \midrule
Decision Tree & 0.4889 & \textbf{0.9778}   & 0.6519                  & 0.3333  & \textbf{1.0}    & 0.5  \\ 
\bottomrule
\end{tabular}
}
\label{table:adaptability}
\end{table}



\begin{table}[h]
\centering
\setlength{\aboverulesep}{0.5pt}
\setlength{\belowrulesep}{0.5pt}
\caption{The table displays an evaluation of cross-column features' effectiveness. Results are compared between models with and without cross-column features.}
{\begin{tabular}{lcccc}
\toprule
       & \multicolumn{4}{c}{Visualization Types (Cross Features)}                \\ \cmidrule{2-5} 
       & \multicolumn{2}{c}{MR}                          & \multicolumn{2}{c}{Hits@2}   \\ \midrule
       & With          & \multicolumn{1}{c|}{Without} & With           & Without \\ \midrule
\tool{} & \textbf{1.626}         &   1.688                        &  \textbf{0.8421}               & 0.8298   \\ \midrule
GQE & \textbf{1.884}     &     2        &    \textbf{0.7268}     &  0.7033      \\ \midrule
 Decision Tree & \textbf{1.893}     &     1.9086          &    \textbf{0.7189}     & 0.7133    \\ \bottomrule
\end{tabular}}\label{table:cross_feature}
\end{table}

\subsubsection{Multiple-Class Visualization Recommendation}

Since each dataset has only the correct visualization type in the previous experiment, the above experiment results fail to demonstrate the adaptability of \tool{}.
Thus, we further conduct experiments in this subsection to evaluate the adaptability of \tool{} in recommending multiple types of visualizations correctly.
A new test set is necessary, where each dataset has more than one correct visualization type.
Since the continuous features of datasets have been transformed into categorical by discretization (Section~\ref{sec: method_KGconstruction}), we grouped datasets with the same feature values. 
The datasets with the same single-column and cross-column features were regarded as a group whose visualization types were interchangeable. In other words, if a dataset is within a group and the group has multiple visualization types, these types will be considered the ground truth for the dataset.
According to observation, a group of datasets with the same feature values are seldom visualized using four visualization types. 
Hence, we tested \tool{} adaptability with the groups of datasets whose visualization types are two and three.
We sampled 406 test datasets with two visualization types and sampled 141 test datasets with three visualization types.



\textbf{Baseline Models.} 
The purpose of the experiment in this subsection is to evaluate the performance of \tool{} in adaptively recommending multiple types of appropriate visualizations.
Given that no existing visualization recommendation approaches are explicitly designed for such a purpose,
we followed the practice of Section~\ref{sec-single-class-vis-recommendation} and also used KG4Vis, GQE, and Decision Tree as the baseline methods
in this experiment. 
\tool{} can suggest adaptive visualization types for an unseen dataset as long as the visualization type entities are within the inference box of \tool{} (Equation \ref{formula: inbox}).
The baseline methods were set to recommend the visualization type with the highest prediction score, which is also common practice when they are used in real applications.


\textbf{Metrics.} For the adaptability experiment, we identify the recommended multiple visualization types based on whether the visualization type entities are inside the box of model inference (Equation~\ref{formula: inbox}). 
Since there are more than one ground truth visualizations for each test dataset,
we used Recall, Precision, and F1 as the metrics in the adaptability experiment~\cite{raschka2014overview}. Recall can evaluate the model's adaptability. A high Recall score means the model can recommend more visualization types suited for a dataset. Precision evaluates the consistency between the model's recommendation results and the ground truth of the dataset. High precision indicates that models are well aligned with users' design preferences. F1 offers a comprehensive measurement that considers both Recall and Precision.

\textbf{Result and Analysis.}
Table~\ref{table:adaptability} shows that \tool{} consistently outperforms the other models in both two and three choices scenarios as indicated by the highest F1 scores, signifying that it effectively balances precision and recall. 
\tool{} also stands out in terms of recall, suggesting that it can recommend a broad range of suitable visualization types and shows excellent adaptability.
In the scenarios with three choices, \tool{}'s performance surpasses all baseline models, further reinforcing its adaptability. Despite some models achieving higher precision in certain scenarios, their lower overall F1 scores imply a lack of diversity in their recommendations.
In contrast, \tool{} maintains high precision across all scenarios,
indicating that its recommendations align well with user selections and can offer a wider range of suitable visualization recommendations.


\subsubsection{Ablation Study about Cross-column Features}
In this ablation study, we evaluated the effects of cross-column features.
Given the crucial role of meaningful relationships between data columns in determining visualization type~\cite{wongsuphasawat2015voyager,wills2010autovis,seo2005rank}, \tool{} incorporates cross-column features.
To verify the significance of our cross-column features, we conducted an ablation experiment. It involves removing the cross-column features from \tool{} and baseline models, and then assessing whether the recommended visualizations still align well with human users' visualization choices.

\textbf{Baseline Models \& Metrics.} 
To evaluate the cross-column features' effect, we used the GQE and Decision Tree as the baseline models. We did not consider the KG4Vis as it does not use the cross-column features. Additionally, the metrics used in the experiment are MR and Hits@2.


\textbf{Result and Analysis.}
Table~\ref{table:cross_feature} displays the effect of cross-column features on the performance of both \tool{} and the baseline models in recommending visualization types.
In each model, i.e., \tool{}, GQE, and Decision Tree, the incorporation of cross-column features resulted in enhanced performance in terms of MR and Hits@2 scores. This is evident when comparing these scores with their counterparts obtained when cross-column features were not used. This trend underscores the importance of integrating cross-column features in the process of visualization recommendation. It suggests that considering the interrelations between columns (i.e., cross-column features), rather than treating each column in isolation, contributes to more effective and relevant visualization recommendations.

\begin{figure*}[!hbt]
    \centering
    \includegraphics[width=1.0\linewidth]{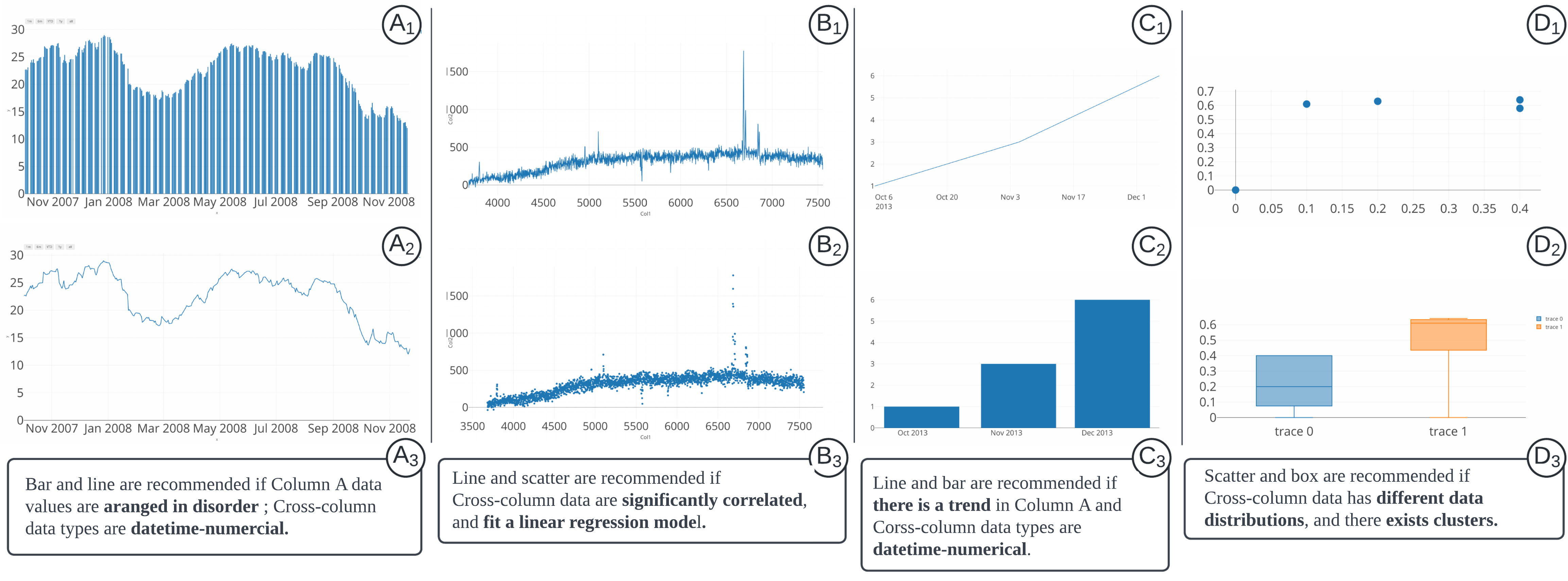}
    \caption{The present figures pairs of visualization recommendations for four different datasets. Visualizations in the same column are from the same dataset. The recommendation results are explained at the bottom of each column. Due to the space limitation, we only describe the top two features in our explanations to illustrate our recommendations. 
    }
    \vspace{-2em}
    \label{fig:multivis}
\end{figure*}

\subsection{Qualitative Evaluation} \label{sec:qualitative} 
To extensively assess \tool{}'s adaptability and understand why these charts are recommended, we further conducted case studies and user interviews to examine recommended visualizations and associated explanations. 

\subsubsection{Adaptability and Explanation}
Figure~\ref{fig:multivis} displays the visualization recommendation results for four datasets by \tool{}. As introduced in Section~\ref{sec:explanation}, \tool{} offers explanations for the recommendation results. 
These explanations highlight the features important to the recommendation results in the understandable language. The following paragraphs describe multiple recommendations for different sets.

Figure~\ref{fig:multivis} \subcomponent{\textbfsub{A}{1}} and  \subcomponent{\textbfsub{A}{2}} show that there are two types of visualizations recommended for a dataset. Figure~\ref{fig:multivis} \subcomponent{\textbfsub{A}{3}} explains the recommendation reason. The two columns' data types are numerical and datetime (values are related to the date or time). Additionally, in the y-axis, the numerical values are not arranged in an orderly manner (e.g., the values of the columns are arranged in increasing or decreasing order). This implies y-axis values are fluctuating. Therefore, the bar and line charts are appropriate for the dataset. The explanation is consistent with existing visualization guidances. For example, Show Me\cite{mackinlay2007show} concludes that a bar chart can be used with two columns of data, one for categorical (datetime can be regarded as categorical value) and another for numerical columns.
Similarly, Munzner~\cite{munzner2014visualization} indicates that a line chart is often used to show temporal trends. 

Figure~\ref{fig:multivis} \subcomponent{\textbfsub{B}{1}} and \subcomponent{\textbfsub{B}{2}} present a line chart and a scatter plot as recommended visualizations for a dataset. Figure~\ref{fig:multivis} \subcomponent{\textbfsub{B}{3}} shows that \tool{} identifies linear correlation
among data columns and further recommends a line chart and a scatter plot to visualize the given dataset. The explanation is also supported by previous work.
According to Cui \textit{et al.}~\cite{cui2019datasite}, a scatter plot is appropriate if there is a correlation between two columns of the dataset, and line charts are recommended if the dataset fits a linear regression model with a low estimated error.

Figure~\ref{fig:multivis} \subcomponent{\textbfsub{C}{1}} and \subcomponent{\textbfsub{C}{2}} show that a line chart and a bar chart are recommended for a dataset by \tool{}. Figure~\ref{fig:multivis} \subcomponent{\textbfsub{C}{3}} gives the explanations for  these two recommendation results: since there is a monotonic trend in the dataset, it is suitable to visualize the dataset using a line chart because the line chart are commonly used to show the trend~\cite{munzner2014visualization}; a bar chart is also recommended for the dataset because it contains categorical and numerical data~\cite{mackinlay2007show}.

Figure~\ref{fig:multivis} \subcomponent{\textbfsub{D}{1}} and \subcomponent{\textbfsub{D}{2}} show that a scatter plot and a box plot are recommended to visualize a dataset. Figure~\ref{fig:multivis} \subcomponent{\textbfsub{D}{3}} reveals important dataset features that led to these recommendation results, i.e., various data distributions and the existence of different clusters.
These explanations also align well with the observations in prior studies.
For example, Cui \textit{et al.}~\cite{cui2019datasite} pointed out that a scatter plot is preferred if several clusters exist in the dataset.
Munzner~\cite{munzner2014visualization} mentioned that a box plot is considered to be appropriate for a dataset with different data distribution across different columns.

These examples show that our explanations for the visualization recommendation align well with the guidelines in prior studies. It confirms the correctness of our visualization recommendation and its adaptability to satisfy the requirements of different datasets.

\begin{figure}[!hbt]
    \centering
    \includegraphics[width=1\linewidth]{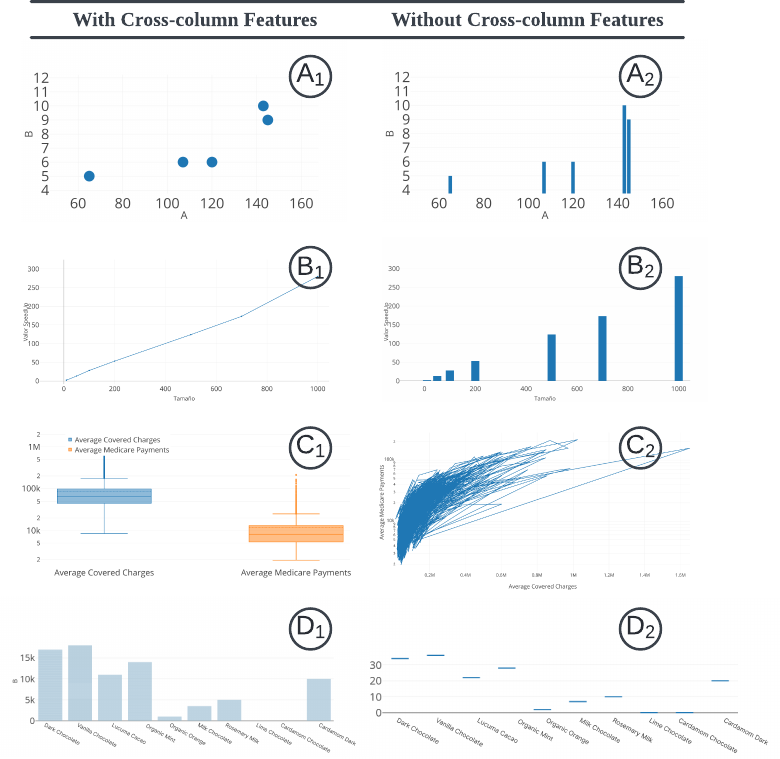}
    \caption{The figure displays pairs of visualization recommendations for four different datasets. A pair of visualization at each row comes from the same dataset. Visualizations recommended by a model with cross-column features are on the left side; Recommendations from the model without cross-column features are on the right side. 
    }
    \vspace{-2em}
    \label{fig:cross-column}
\end{figure}


\subsubsection{Cross-column Features}

We compared recommendations generated by \tool{} with and without cross-column features to demonstrate their necessity.
Several recommended visualizations are shown in Figure~\ref{fig:cross-column}, where the visualizations on the left are recommended by considering cross-column features, and those on the right are recommended without considering cross-column features.
To explore how cross-column features lead to different recommendations, we studied the explanations regarding cross-column features. 

Figure~\ref{fig:cross-column}~\subcomponent{\textbfsub{A}{1}} and \subcomponent{\textbfsub{A}{2}} show that for the same dataset, \tool{} with cross-column features recommended scatter plot. In contrast, \tool{} without cross-column features recommended a bar chart. 
According to the model's explanation, \textit{``there are clusters between two data columns and columns' data types are numerical-numerical"}. 
Since apparent clusters among data columns and two data columns are both numerical, the scatter plot is appropriate. 
In contrast, the bar chart is not suitable for
a dataset consisting of two numerical data columns~\cite{mackinlay2007show}. Therefore, the inappropriate visualization type is recommended because it does consider the cross-column features.

As shown in Figure~\ref{fig:cross-column}~(\subcomponent{\textbfsub{B}{1}}, \subcomponent{\textbfsub{B}{2}}), \tool{} with cross-column features recommended a line chart rather than a bar chart because \textit{``two column data values are significantly correlated, and data types are numerical-numerical"}. The study by Saket~\textit{et al.}~\cite{saket2018task} indicates that a line chart is suitable for finding correlation. \tool{} with cross-column features capture the critical interrelationships between data columns in a dataset and thus recommend a line chart.

 Figure~\ref{fig:cross-column}~\subcomponent{\textbfsub{C}{1}} and \subcomponent{\textbfsub{C}{2}} present a box plot recommended by \tool{} with cross-column features and line chart without cross-column features. Based on the corresponding explanation, we recognized that  \textit{``Cross-column has a different distribution"}. That important cross-column feature led to the box plot recommendation.

 Figure~\ref{fig:cross-column}~\subcomponent{\textbfsub{D}{1}} and \subcomponent{\textbfsub{D}{2}} show a bar chart and a box plot, respectively.
 Based on the explanation, the important cross-column feature in the recommendation is that \textit{``two column data types are categorical-numerical''}, and thus recommended a bar chart rather than a box plot. As established by Mackinaly \textit{et al.}~\cite{mackinlay2007show}, a bar chart is well-suited for visualizing data involving two columns with categorical and numerical values. Therefore, a bar chart is an appropriate choice for visualizing the dataset. Additionally, this case illustrates that without cross-column features, \tool{} does not take the data type of two columns into account, leading to a box plot recommendation that violates the visualization guidance.
 

These cases demonstrate the importance of cross-column features in visualization recommendations. 
According to our observations, cross-column features can help \tool{} to exclude recommendations that violate visual design rules restricted by cross-column features. 
Cross-column features can reveal interrelationships between columns.
Therefore, it is necessary to consider cross-column features in 
recommending appropriate visualizations.

\subsubsection{User Interview and Feedback}\label{sec: user_interview}
Our quantitative experiments above demonstrate the effectiveness of \tool{} in adaptively recommending multiple appropriate visualizations, but it is also crucial to ask actual users to evaluate the recommended visualizations as well as the natural language explanations provided by \tool{}.
Thus, we conducted user interviews with two distinct tasks, where each is designed to evaluate one aspect of
\tool{}:
the adaptability of our recommendations and the clarity of the generated explanations:
\begin{enumerate}
    \item[\textbf{Task 1}]\textbf{Recommendation Adaptability Assessment.} Participants were asked to view the recommended visualizations for ten randomly sampled datasets with multiple recommended visualization options. Provided with the tabular dataset, participants needed to assess and provide feedback on how well each recommended visualization presented the original tabular data of the dataset.
    \item[\textbf{Task 2}]\textbf{Explanation Clarity Assessment.} Participants were provided with explanations for the recommendations of ten randomly sampled datasets. They were then asked to evaluate whether the explanations helped them understand the recommendation results. Feedback was requested on the clarity of the explanations.
    \end{enumerate}

For the user interviews, we recruited 12 participants, all of whom actively use data visualization tools but have varying degrees of expertise in the field. This allowed us to conduct a thorough evaluation of \tool{} across a range of user experiences. For our analysis, we categorized them into two groups. The first group (E1-E6) is composed of participants who have demonstrated a high level of expertise in data visualization, evidenced by their contributions to at least one scientific publication in the field. The second group (C1-C6) includes participants who regularly use data analysis tools, such as Excel, and have a fundamental understanding of data science.
The diversity of participants' backgrounds
allowed us to assess 
the
effectiveness of \tool{} from different perspectives.
Throughout the interview, we encouraged participants to express their thoughts and feedback in a think-aloud manner.


After finishing the interviews, we analyzed all the feedback from participants and also categorized participant feedback into two groups accordingly.
We then conducted a thematic analysis~\cite{guest2011applied} within each group to identify recurrently-raised issues. To highlight the differences between the two groups' feedback, we conducted a cross-comparison of these issues. Consequently, our analysis of the feedback revealed both convergences and divergences from the perspectives of data visualization experts and common users, which are organized and presented as follows:

\textbf{Recommendations of Multiple Visualization Types.} Overall, 
all participants found that most of the visualization recommendations by \tool{} are appropriate for the given datasets. 
For instance, E3 endorsed the variety in our recommendations: \textit{``It is reasonable to recommend these multiple types of visualizations for the same dataset. 
For instance, when examining a trend or investigating correlations and distributions within a dataset, 
line chart, bar chart, and scatter plot can all be used to visualize the same dataset".} 
However, we also observed a discrepancy between these two participant groups. The second group's (C1-C6) acceptance rate of visualization recommendations seems to be influenced by their existing knowledge of data visualization, while data visualization experts are not. For instance, C1 
has never used a box plot before, and she
got confused when she was presented with a box plot as the visualization recommendations. Similarly, C6 disagreed with some of the recommended line charts for datasets without a column being the time variable, as she insists that 
line charts should be predominantly used for time-series-related data.
A significant suggestion from both groups was to further incorporate the users' analytical tasks when recommending appropriate types of visualizations for a given dataset.

\textbf{Recommendation Explanations.} The clarity of our explanations was confirmed by all the participants regardless of their familiarity with data visualizations. 
They pointed out that these explanations can help them effectively and conveniently understand why specific visualizations were recommended for a given dataset.
For example, C6 mentioned that the explanations enhanced her comprehension of the recommended visualization, as they highlighted the specific features that drove the recommendations.
Nevertheless, we observed that a user's knowledge level of visualization significantly influences their requirements for explanations. Users less familiar with data visualization may need more detailed and contextual explanations of the recommended visualization. For example, C4 suggested that the explanation should be task-oriented, displaying a specific scenario, and C2 expressed a desire for justifications in explanation when his unfamiliar visualizations are recommended. Conversely, users with more advanced knowledge, like E2, might prefer an explanation that focuses on the characteristics of a dataset. This observation underscores the importance of tailoring explanations to the knowledge level and needs of individual users to facilitate their understanding and acceptance of the recommended visualizations.
Also, participants have provided insightful suggestions for further enhancing \tool{}.
For instance,
a common suggestion from both groups of participants was that our explanation should also illustrate why certain types of visualizations were not recommended beyond only explaining why certain visualizations were recommended.
More specifically, E1
advised that \tool{} could incorporate a \textit{What-if} functionality, enabling users to discern which features are most influential to recommendation results.
Moreover,
some common users (C6, C5)
pointed out that certain terms used in the explanation, such as 
``disorder",
should be presented in a more intuitive manner.

%% file: src/5-discussion.tex
\section{Discussion}

\subsection{Lessons}

\textbf{Explanability.} 
Our explanation for visualization recommendations considers both feature importance and intuitiveness.
120 features are used in our approach to comprehensively model the characteristics of the input dataset.
However, an increasing number of dataset features can also result in the difficulty of providing intuitive explanations for the visualization recommendation from the perspective of dataset features.
To strike a good balance between visualization recommendation performance and explainability, we integrated an attention mechanism in \tool{} (Section~\ref{sec:explanation}) that can identify critical features for the final visualization recommendation result.
Our explanation is built upon the most important two or three features.
For feature intuitiveness, some dataset features are important for the final visualization recommendation result, but their meanings are difficult to interpret,
especially for statistical data features (i.e., the entropy is high, 
Kolmogorov–Smirnov test result is significant). Given that target audiences comprise common users, these perplexing features
were avoided in the final
explanations.

\textbf{Trade-off in Model Training Strategy.} \tool{} performs two kinds of recommendation tasks: axis and visualization type recommendation. Since single-column features are required in both axes and visualization type prediction, the embeddings of the single-column features are influenced by two tasks about axis and visualization types recommendation, when we combine the two tasks for the model to learn; the multi-task learning can introduce noise because visualization type recommendation visualization is unrelated to axes prediction. To investigate the effect, we conducted a control experiment on multi-task and single-task learning; the model was doing multi-task learning while two separate models were trained for each task in the control group. According to the experiment result, training different models for each task led to a few points of improvement. However, the improvement came at the cost of double the computation time and storage, because it needs to train two individual models to do axis and visualization type recommendation task, respectively. 

\textbf{Feature Importance.} 
\tool{} defines a comprehensive set of features, including 80 single-column features and 40 cross-column features, to capture the diverse characteristics of input datasets.
A feature importance analysis revealed that some single-column features, especially those related to column names and data column statistical properties, are particularly impactful.
Among these, \textit{data column length} emerges as a highly influential feature. The significance of data column length can be attributed to its role in reflecting data density, which in turn informs the choice of visualization type. For instance, for a high-density dataset,
a line chart may be
more suitable
than a bar chart or scatter plot, as it represents data points in a less cluttered and more understandable way.
Also, our analysis highlights the importance of \textit{digits in column name}. This feature carries semantic information about the data column, which potentially indicates users' design logic for visualization. For example, a column named "Year2017" can imply the need to visualize a trend over time, while column names such as "Method1" or "Method2" only reveal the need for a comparison.
\subsection{Limitations}

\alexin{\textbf{Corpus.}
We utilize the dataset-visualization pairs uploaded by Plotly users as the ground truth, and most visualization choices of them align well with general visualization design guidelines. However, there are also a small number of problematic visualization choices for the input datasets, which may have a negative impact on the performance of \tool{}.
To further bolster the performance of \tool{}, it is important to expand our corpus with more high-quality dataset-visualization pairs. For example, we can try to collect more data visualization examples 
created by experienced visualization experts 
from professional visualization blogs or forums like Observable\footnote{https://observablehq.com/}.
}

\alexin{\textbf{Visualization Choices.}
As an initial step towards adaptive and explainable visualization recommendation,
\tool{} is applied to
the widely-used standard charts in this paper,
like line charts, bar charts, scatter plots, and box plots, and it does not encompass all types of visualizations.
Such a 
choice
originates from both the popularity of these visualizations~\cite{hu2019vizml} and the fact that other visualizations are scarce in the Plotly corpus.
Also, given that these standard charts have an emphasis on the axes' visual encodings~\cite{hu2019vizml}, \tool{} mainly focuses on recommending appropriate types of data visualizations and the x/y axes. 
However, with a new dataset-visualization corpus of other types of visualizations, \tool{} can be easily extended to work for new types of visualizations and other detailed visualization encodings like color schemes.
}

\textbf{User-centric Recommendation.}
\tool{} effectively maps datasets to visualizations but lacks an explicit consideration of users' specific intents, such as their analytical tasks or preferences.
As indicated by user feedback (Section~\ref{sec: user_interview}),
it will be interesting to further incorporate user intent in visualization recommendations, which can ensure that the recommended visualizations align with the user's specific needs. 
In this paper, we demonstrate the effectiveness of \tool{} by using datasets with two columns,
but \tool{} can also
recommend appropriate visualizations for datasets with more than two columns.
It can be achieved by extracting cross-column features from every possible combination of two columns in the dataset, and
then finding the intersection of these cross-column box embeddings, which indicate the dataset's characteristics and can be further used to derive the appropriate visualization choices.
The possible issue of extending \tool{} to datasets with over two columns is that the exhaustive search of every possible combination of two columns in the dataset can be time-consuming, which can be mitigated by further considering user intent to narrow down the search space of pairwise column combination in the visualization recommendation process.
\alexin{In addition, some terminologies used in the natural language explanations by \tool{}
may not
be easily understood by
all users. For instance, technical terms like “a linear regression model” are obvious for machine learning practitioners
but can be perplexing for laypersons. 
It will be helpful to further incorporate more straightforward explanations into \tool{}, making it more accessible to a broader range of audiences. }

\textbf{Training Time.}
The training time of \tool{} is prolonged due to a large number of extracted features and corpus.
\tool{} extracts many features from a dataset. These features enable \tool{} to comprehensively model the dataset characteristics and further increase the generability of \tool{}, which can recommend appropriate visualization types for diverse datasets. However, the large number of features also increases model complexity and thus enlarges the model size. In addition, to better learn the complex mapping from datasets to visualization types, \tool{} is trained on a large corpus. Feature importance analysis can be used to identify unimportant features and reduce the feature number in the model.

%% file: src/6-conclusion.tex
\section{Conclusion and Future Work}
In this paper, we propose \tool{}, an adaptive and explainable approach for visualization recommendation. 
Given a dataset, \tool{} 
can adaptively recommend
multiple
appropriate visualization choices 
and
provide
detailed explanations for the recommendation result. Our approach consists of four modules: feature extraction, knowledge graph construction, model training, and inference.
It first extracts
the individual column's features and the interrelationship among data features, data columns and visualization choices. With these features and interrelationships, 
a knowledge graph is constructed to model them. The box embeddings of entities and relations in the knowledge graph can also be learned.
With these learned box embeddings, 
an inference module can adaptively recommend multiple visualizations for an unseen dataset and provide natural language explanations for the recommendations.
Quantitative and qualitative evaluations are conducted to evaluate the effectiveness and adaptability of \tool{}.

In future work, we will collect more diverse dataset-visualization pairs and extend \tool{} to recommend more different types of data visualizations in an adaptive and explainable manner. Also, it is interesting to investigate how user intent can be integrated into \tool{} to further improve its efficiency and effectiveness in adaptive and explainable visualization recommendations.

%% file: src/appendix.tex
\setcounter{table}{0}
\renewcommand{\thetable}{A\arabic{table}}
\onecolumn

\appendix
\renewcommand{\thesubsection}{\Alph{subsection}}

This section introduces the detailed features used in \tool{}. We extract 120 features from the dataset, where 80 are single-column features, and 40 are cross-columns features. Table~\ref{table:feature_list} displays all single-column features and corresponding meanings, and Table~\ref{table:cross_feature_list} shows cross-column features and their meanings.
\setlength{\belowrulesep}{1.0 pt}
\begin{longtable}{l|l}
    \caption{The table shows single-column features and their explanations.}
    \label{table:feature_list} \\   



\toprule
Feature Name                                                                                                             & Explanation                                                                                                           \\ 
\hline
num\_unique\_elements                                                                                                                                                                                    & The number of unique values in a  data column.                                                                             \\ 
\hline
sortedness                                                                                                                                                                                               & The quality of being sorted of values in a data column.~                                                                     \\ 
\hline
unique\_percent                                                                                                                                                                                          & The percentage of unique value in a data column.                                                                     \\ 
\hline
\begin{tabular}[c]{@{}l@{}}percent\_outliers\_15iqr\\percent\_outliers\_3iqr\\percent\_outliers\_3std\\percent\_outliers\_1\_99\end{tabular}                                                                                 & They measure the proportion of outliers  according to 1.5IQR/3IQR/3Std/(1\%, 99\%).                                       \\ 
\hline
\begin{tabular}[c]{@{}l@{}}entropy\\gini\end{tabular}                                                                                                                                                    & A data column's disorderliness.                                                                                            \\ 
\hline
\begin{tabular}[c]{@{}l@{}}skewness\\kurtosis\\moment\_5\\moment\_6\\moment\_7\\moment\_8\\moment\_9\\moment\_10\end{tabular}                                                                            & Distribution characteristics of values for a data column.                                                                  \\ 
\hline
\begin{tabular}[c]{@{}l@{}}lin\_space\_seq\_coef\\log\_space\_seq\_coef\end{tabular}                                                                                                                     & The degree of a data column values in the linear/ logarithmic range.                                                           \\ 
\hline
\begin{tabular}[c]{@{}l@{}}quant\_coeff\_disp\\med\_abs\_dev\\avg\_abs\_dev\\coeff\_var\\std\\var\end{tabular}                                                                                           & Variation characteristics of values for a data column.                                                                     \\ 
\hline
\begin{tabular}[c]{@{}l@{}}normality\_p\\normality\_statistic\end{tabular}                                                                                                                               & The degree of a data column values is in the normal distribution.                                                          \\ 
\hline
\begin{tabular}[c]{@{}l@{}}normalized\_range\\range\end{tabular}                                                                                                                                         & The range of a data column values.                                                                                          \\ 
\hline
\begin{tabular}[c]{@{}l@{}}q25\\q75\\normalized\_median\\normalized\_mean\\min\\max\\mean\\median\\length\end{tabular}                                                                                   & A description of a data column's statistical characteristics.                                                              \\ 
\hline
percent\_of\_mode                                                                                                                                                                                            & The percente of mode values in a data column.                                                                              \\ 
\hline
\begin{tabular}[c]{@{}l@{}}num\_none\\percentage\_none\end{tabular}                                                                                                                          & Numer/ percentage of missing values in a data column.                                                                      \\ 
\hline
\begin{tabular}[c]{@{}l@{}}mean\_value\_length\\median\_value\_length\\min\_length\_of\_value\\std\_length\_of\_value\\max\_length\_of\_value\end{tabular}                                                                                                                       & length of mean/ median/ minimum/ standard deviation/ maximum values in a data column.                                                                             \\ 
\hline
has\_none                                                                                                                                                                                                & The data column has missing values.                                                                                        \\ 
\hline
\begin{tabular}[c]{@{}l@{}}is\_monotonic\\is\_sorted\\is\_unique\end{tabular}                                                                                                                            & Data column values are montonic/sorted/unique.                                                                             \\ 
\hline
\begin{tabular}[c]{@{}l@{}}is\_lin\_space\\is\_log\_space\end{tabular}                                                                                                                                   & Data column values in linear/ logarithmic range.                                                                           \\ 
\hline
\begin{tabular}[c]{@{}l@{}}has\_outliers\_15iqr\\has\_outliers\_3iqr\\had\_outliers\_3td\\has\_outlisers\_1\_99\end{tabular}                                                                             & Outliers exists according to 1.5IQR/ 3IQR/ 3Std/ (1\%,99\%) rule.                                                          \\ 
\hline
\begin{tabular}[c]{@{}l@{}}is\_normal\_1\\is\_normal\_5\end{tabular}                                                                                                                                     & The data column values are significantly in the normal distribution with \textit{p~}\textless{} 0.01/ p \textless{} 0.05.  \\ 
\hline
\begin{tabular}[c]{@{}l@{}}number\_of\_words\_in\_name\\number\_of\_uppercase\_char\end{tabular}                                                                                                                     & The number of words/ upper case characters in a data column name.~                                                         \\ 
\hline
name\_length                                                                                                                                                                                             & The length of a data column name.                                                                                          \\ 
\hline
field\_name\_length                                                                                                                                                                                      & The number of characters in a data column name.                                                                            \\ 
\hline
\begin{tabular}[c]{@{}l@{}}data\_type\_is\_string\\data\_type\_is\_integer\\data\_type\_is\_decimal\\data\_type\_is\_datetime\end{tabular}                                                                                      & The data column value type is string/ integer/ decimal/ datatime.                                                          \\ 
\hline
\begin{tabular}[c]{@{}l@{}}general\_type\_is\_t\\general\_type\_is\_q\\general\_type\_is\_c\end{tabular}                                                                                                    & The general type of a data column values is temporal/ quantitative/ categorical.                                          \\ 
\hline
first\_char\_upper\_name                                                                                                                                                                                           & The data column name starts with an upper case character.                                                                  \\ 
\hline
\begin{tabular}[c]{@{}l@{}}x\_in\_name\\y\_in\_name\\id\_in\_name\\time\_in\_name\\digit\_in\_name\\space\_in\_name\\dollar\_in\_name\\pounds\_in\_name\\euro\_in\_name\\yen\_in\_name\end{tabular} & \begin{tabular}[c]{@{}l@{}}There exist special characters: ``x", ``y", ``id", time, digit, whitespace, `\$", ``\pounds", ``\texteuro", and ``\yen" in a \\column name.\end{tabular}           \\
\bottomrule
\end{longtable}

\setlength{\belowrulesep}{1.0 pt}
\setlength{\aboverulesep}{1.0 pt}
\begin{longtable}{l|l}
    \caption{The table shows cross-column features and their explanations.}
    \label{table:cross_feature_list} \\   
                                        
\toprule
Feature Name                                                                                                             & Explanation                                                                                                           \\ 
\hline
\begin{tabular}[c]{@{}l@{}}percent\_shared\_elements\\percent\_shared\_unique\_elements\end{tabular}                                                                                 & The percentage of values/ unique values shared by pairwise columns.                        \\ 
\hline
\begin{tabular}[c]{@{}l@{}}num\_shared\_element\\snum\_shared\_unique\_elements\end{tabular}                                                                                         & The number of values/ unique values shared by pairwise columns.                            \\ 
\hline
num\_shared\_words                                                                                                                                                                   & The number of words that are shared by two column names.                                   \\ 
\hline
percent\_shared\_words                                                                                                                                                               & The percentage of words that are shared by two column names.                               \\ 
\hline
percent\_range\_overlap                                                                                                                                                              & The percentage of the overlapping value range in two data columns.                         \\ 
\hline
has\_range\_overlap                                                                                                                                                                  & Two columns have overlapping value ranges.                                                 \\ 
\hline
\begin{tabular}[c]{@{}l@{}}has\_shared\_elements\\has\_shared\_unique\_elements\end{tabular}                                                                                         & There exist some values/ unique values in two data columns.                                \\ 
\hline
has\_shared\_words                                                                                                                                                                   & There exist some words in two data column names.                                           \\ 
\hline
\begin{tabular}[c]{@{}l@{}}identical\\identical\_unique\end{tabular}                                                                                                                 & Two data column values/ unique values are identical.                                       \\ 
\hline
linregress\_err                                                                                                                                                                      & The standard error of the calculated linear regression for two data columns.               \\ 
\hline
linregress\_p                                                                                                                                                                        & The probability that there exists a linear regression.                                     \\ 
\hline
\begin{tabular}[c]{@{}l@{}}kmeans\_3\_avg\_err\\kmeans\_5\_avg\_err\\kemans\_6\_avg\_err\end{tabular}                                                                                & The average distance between the data of two columns and calculated 3/ 5/ 6 centroids.     \\ 
\hline
correlation\_value                                                                                                                                                                   & The value of the correlation coefficient between two columns of data.                      \\ 
\hline
correlation\_p                                                                                                                                                                       & The probability that there exists a correlation between two columns.                       \\ 
\hline
\begin{tabular}[c]{@{}l@{}}ks\_p\\chi2\_p\\one\_way\_anova\_p\end{tabular}                                                                                                           & The probability that these two columns fit common statistical hypotheses.                  \\ 
\hline
\begin{tabular}[c]{@{}l@{}}ks\_statistics\\chi2\_statistic\\one\_way\_anova\_statistic\end{tabular}                                                                                  & The calculated value between two data columns with common statistical testing approaches.  \\ 
\hline
edit\_distance                                                                                                                                                                       & The dissimilarity between two data column names.                                           \\ 
\hline
normalized\_edit\_distance                                                                                                                                                           & The normalized dissimilarity between two data column names.                                \\ 
\hline
nestedness                                                                                                                                                                           & The degree of two data column values are nested.                                            \\ 
\hline
\begin{tabular}[c]{@{}l@{}}chi2\_significant\_005\\correlation\_significant\_005\\ks\_significant\_005\\linregress\_significant\_005\\one\_way\_anova\_significant\_005\end{tabular} & Two data columns fit common statistical hypotheses with a p-value = 0.05.                  \\ 
\hline
\begin{tabular}[c]{@{}l@{}}categorical\_categorical\\category\_numerical\\numerical\_numerical\\time\_categorical\\time\_numerical\\time\_time\end{tabular}                          & The relationship between data types of two columns.                                         \\
\bottomrule
\end{longtable}